\documentclass[twocolumn,showpacs,preprintnumbers,amsmath,amssymb]{revtex4}
\usepackage[dvips]{graphicx}
\usepackage{dcolumn}

\begin{document}

\title
{Half-Quantum Vortex Pair in Polar-distorted B Phase of Superfluid $^3$He in Aerogels
}

\author{Masaki Tange and Ryusuke Ikeda}

\affiliation{Department of Physics, Graduate School of Science, Kyoto University, Kyoto 606-8502, Japan
}

\date{\today}

\begin{abstract} 
Motivated by the recent observation and argument on a large half-quantum vortex (HQV) pair connencted by a Kibble wall in superfluid $^3$He in nematic aerogels, we numerically study to what extent a huge HQV pair can intrinsically occur with no pinning effect due to the aerogel structure in the polar-distorted B (PdB) phase of superfluid $^3$He. By fully examining the impurity-scattering induced pairing vertex, the emergence of Anderson's Theorem in the $p$-wave superfluid is verified in the two opposite limits, the isotropic and strongly anisotropic limits. Solving numerically the resulting Ginzburg-Landau (GL) free energy in the weak-coupling approximation and by taking account of the Fermi-liquid (FL) corrected gradient terms, the anisotropy dependence of the vortex structure minimizing the free energy is examined. It is found that, close to the transition between the polar and PdB phases, an interplay of the strong anisotropy and the FL correction makes emergence of a large HQV pair in the PdB phase possible, and that, nevertheless, such a large pair easily shrinks deep in the PdB phase, indicating that a pinning effect due to the aerogel structure is necessary in order to keep a large pair size there. The obtained result indicates the validity of the London limit for describing the vortex structure, and a consistency with the picture based on the NMR measurement is discussed. 
\end{abstract}



\maketitle

\section{Introduction}
The recent observation of half-quantum vortices (HQVs) \cite{Autti} in the polar superfluid phase, proposed to appear in superfluid $^3$He in anisotropic aerogels through a model calculation \cite{AI06} assuming a weak anisotropy and experimentally discovered in nematic aerogels \cite{VVD}, has opened a new door for studying possible vortices in a fermionic superfluid phase. A nematic aerogel has its strands aligned to one direction and can be regarded, broadly speaking, as a collection of line-like obstacles. The HQV has originally been expected to be realized in the thin film configuration of the chiral superfuid A phase with its orbital angular momentum locked perpendicularly to the film plane \cite{VM76,SV85}. However, the chiral A phase is realized with the help of the strong-coupling correction which is effective at higher pressures, while it has been clarified \cite{NI1} that the HQV tends to be destabilized by the strong coupling correction. Fortunately, the polar phase realized in the nematic aerogel has a wider temperature range of its stability at relatively lower pressures, and hence, a superfluid $^3$He in the nematic aerogels becomes the best playground for studying this novel topological object. 

Recently, experimental investigation on the vortices in the nematic aerogel has been extended to lower temperatures \cite{Eltsov}, and the HQVs have been found through the NMR measurements to survive in the A pand B phases realized at lower temperatures in the nematic aerogels. Since such A and B phases in the nematic aerogel are distorted by the anisotropy of the scattering events due to the aerogel structure, the resulting A and B phases will be called hereafter as the polar-distorted A (PdA) and PdB phases following Ref.\cite{Eltsov}, It has been suggested that the detected HQV-pairs do not change their positions upon both the cooling from and the warming to the polar phase, and hence that, since in their rotated experiments the rotation axis is parallel to the direction to which the strands are aligned, realization of such surprising events is largely supported by a strong pinning effect due to the line-like aerogel structure \cite{Eltsov}. There, however, just the method of analyzing the NMR data on the basis of a hypothetical description of the vortex structure in the London limit has been presented, and the validity of their London description has not been examined. We note that a similar anisotropic growth of the half-core structure of the double-core vortex occurring upon cooling in the context of the superfluid $^3$He in {\it isotropic} aerogels cannot be described based on the London limit \cite{NI2}. Therefore, it is natural to ask to what extent the huge HQV pairs realized in the PdA and PdB phases are intrinsically stable and whether the description in the London limit is justified or not. 

In the present work, we start with reformulating the Ginzburg-Landau (GL) approach for describing the superfluid $^3$He in anisotropic aerogels by extending the weakly anisotropic model \cite{AI06} of the impurity-scattering potential to the strongly anisotropic case appropriate for the situations in the nematic aerogel \cite{Autti,VVD,VVD2}. Using the resulting GL free energy, stable vortex solutions are studied in both the polar and PdB phases in strongly anisotropic cases. Throughout this work, we focus on the weak-coupling approximation neglecting the strong-coupling correction to the bulk free energy terms because incorporating the strong-coupling correction in the strongly anisotropic case has not been formulated so far. For this reason, the PdA phase never appears in the present results, and we have only a direct continuous transition between the polar and PdB pairing states. Since our model covers, in the weak-coupling approximation, the well-known isotropic case which has essentially the same vortex solution as in the bulk liquid case, we study how the nonaxisymmetric double-core vortex is changed and stabilized with increasing the anisotropy on the impurity-scattering process. The core structure of the double-core vortex is often called as a half-core pair, because, under an appropriate condition, the description of the half-core pair based on the London limit, i.e., a HQV pair connected by a planar wall \cite{Volovik90} becomes appropriate. Throughout this work, the double-core vortex will be identified with a HQV pair {\it only when the planar wall is well defined and clearly visible}. It is found that, as the anisotropy is increased, the description in the London limit of the order parameter profiles of one HQV-pair becomes better. Further, as the anisotropy is 
increased, 
the separation between the two HQVs forming one pair is increased and, in particular, becomes macroscopic close to the transition temperature $T_{\rm PB}$ between the polar and PdB phases. However, this size rapidly shrinks upon cooling from $T_{\rm PB}$, accompanying an increase of the tension of the Kibble wall of the polar-distorted planar state \cite{Volovik19} upon cooling. It implies that, deep in the PdB phase, a HQV pair with a macroscopic size is not naturally stabilized and hence, justifies the picture \cite{Eltsov} of a HQV pair stabilized by the pinning due to the line-like aerogel structure. 

The present paper is organized as follows. In sec.II, the vertex correction to the pairing process is explained in details together with the model of the impurity scattering used in this work. In sec.III, the resulting GL free energy affected by the impurity effects is explained. In sec.IV, it is explained how a HQV pair in the PdB phase is stabilized within the description in the London limit. Our numerical results and detailed discussions about them are presented in sec.V, and a summary and discussions are given in sec.VI. Details on the impurity-induced vertex correction and its effects on the O($|\Delta|^4$) gradient terms are explained in two Appendices.

\section{Model of Impurity Scattering}

Our microscopic analysis for deriving the GL free energy is based on a BCS Hamiltonian with the nonmagnetic and random scattering potential term of the form 
\begin{equation}
{\cal H}_{\rm imp} = \int_{\bf r} \psi^\dagger_\sigma({\bf r}) u({\bf r}) \psi_\sigma({\bf r}). 
\end{equation}
Here we focus on the case in which the scattering process is nonmagnetic, since the local surface of the aerogel is implicitly assumed to be entirely coated by $^4$He so that the spin-flip scattering between the solid $^3$He on the surface and the quasiparticles of the liquid $^3$He is ineffective. Regarding the random averaging over $u({\bf r})$, the Fourier transform $u({\bf k})$ of $u({\bf r})$ is assumed to have zero mean and the mean-squared average 
\begin{equation}
{\overline {|u_{\bf k}|^2}} = \frac{1}{2 \pi N(0) \tau} w({\bf k}). 
\end{equation}

In the original work \cite{AI06}, a weak anisotropy has been incorporated in $w({\bf k})$ in the form 
\begin{equation}
w({\bf k}) = 1 - \delta {\hat k}_z^2, 
\end{equation}
where ${\hat k}_z = k_z/k_{\rm F}$. The positive constant $\delta$ in eq.(3) corresponds to $-\delta_u$ in Ref.\cite{AI06} where a narrow polar phase has been proposed to appear in an aerogel sample stretched along the $z$-direction. For this reason, the $z$-axis will be called as the polar axis. Hereafter, as an extention of this impurity-scattering model to the case in a strongly anisotropic aerogel, the following model on $w({\bf k})$ will be used: 
\begin{equation}
w({\bf k}) = \frac{1 + (\sqrt{\delta} - 1)\theta(\delta - 1)}{ 1 + \delta {\hat k}_z^2}. 
\end{equation}
In the weakly anisotropic limit where $\delta \ll 1$, this expression reduces to eq.(3), while, in the opposite strongly anisotropic limit where $\delta \to +\infty$, eq.(4) approaches the quantity 
\begin{equation}
w_\infty({\bf k}) = \pi k_{\rm F} \delta(k_z). 
\end{equation}
The factor $\sqrt{\delta}$ in eq.(4) in $\delta \geq 1$ is necessary to obtain a physically reasonable limit of the quasiparticle relaxation rate in $\delta \to \infty$. Equation (5) implies that, in the limit of strong anisotropy, the scattering event is specular along the polar axis. This corresponds to the model proposed by Fomin \cite{Fomin18} regarding the nematic aerogels as a collection of columnar defects.  

\begin{figure}[t]
{
\includegraphics[scale = 0.6]{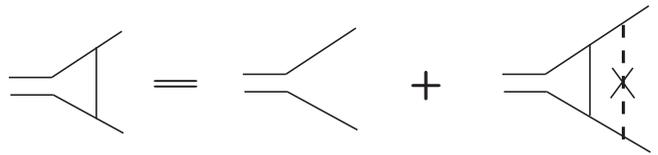}
}
\caption{Diagram expressing the Bethe-Salpeter equation the vertex function $\Lambda$ (triangle) obeys. The straight solid lines mean normal Green's functions, and the impurity average of the squared random potential is expressed by a dashed line with a cross. 
}
\label{s:fig:Vertex}
\end{figure}
To derive a GL free energy incorporating the scattering processes via the anisotropic aerogel structure with any strength of the anisotropy, we need a renormalized pairing vertex $\Lambda_j$ replacing the bare pairing vertex ${\hat p}_j = p_j/k_{\rm F}$, which depends not only on the relative momentum ${\bf p}$ but also on the center-of-mass momentum ${\bf k}$ and the fermion Matsubara frequency $\varepsilon$. The Bethe-Salpeter equation sketched in Fig.1 
\begin{eqnarray}
\Lambda_j({\hat {\bf p}}; {\bf k}) &=& {\hat p}_j + \frac{1}{2 \pi N(0) \tau} \int_{{\bf p}'} \Lambda_j({\hat {\bf p}'}; {\bf k}) \nonumber \\ 
&\times& {\cal G}_\varepsilon({\bf p}'+{\bf k}/2) {\cal G}_{-\varepsilon}(-{\bf p}'+{\bf k}/2) w({\bf p} - {\bf p}')
\end{eqnarray}
for $\Lambda_j$ can be solved in a closed form by assuming $\Lambda_j$ to take 
the expression
\begin{eqnarray}
\Lambda_j({\hat {\bf p}}; {\bf k}) &=& {\hat p}_j ( \delta^{({\rm z})}_{ij} + {\hat z}_j ({\hat z}_i C({\bf k}) + v k_i v k_z C_{1z}) \, ) \nonumber \\
&-& {\rm i} s_\varepsilon v k_j ( \, \delta^{({\rm z})}_{ij} B({\hat p}_z) + {\hat z}_i {\hat z}_j ( \, B({\hat p}_z) + D({\hat p}_z) \, ) \, ), 
\end{eqnarray}
where 
\begin{equation}
{\cal G}_\varepsilon({\bf p}) = ({\rm i}{\tilde \varepsilon}_{\bf p} - \xi_{\bf p})^{-1}
\end{equation}
is the Matsubara Green's function with the self energy term including the 
impurity scattering, $\xi_{\bf p}$ is the quasiparticle energy, and ${\tilde \varepsilon}_{\bf p}$ will be given by eq.(31) in Appendix. Further, 
$\delta^{({\rm z})}_{ij} = \delta_{ij} - {\hat z}_i {\hat z}_j$, $s_\varepsilon = \varepsilon/|\varepsilon|$, $v {\hat p}_j$ is the Fermi velocity, and 
\begin{eqnarray}
C({\bf k}) &=& C_0(\varepsilon) + C_{21}(\varepsilon) v^2 k^2 + C_{2z}(\varepsilon) v^2 k_z^2, \nonumber \\
B({\hat p}_z) &=& B_0(\varepsilon) + \Delta B(\varepsilon) {\hat p}_z^2, \nonumber \\
D({\hat p}_z) &=& D_0(\varepsilon) + \Delta D(\varepsilon) {\hat p}_z^2. 
\end{eqnarray}

The general form of $\Lambda_j$ is involved and will be presented in Appendix. Here, just its limiting expressions will be explained. 
First, in the isotropic limit where $\delta \to 0$, $C_0 \to 1$, and other coefficients except $B_0$ vanish. Then, 
$\Lambda_j$ reduces to 
\begin{equation}
\Lambda^{(0)}_j = \delta_{ij} \biggl({\hat p}_j - {\rm i} s_\varepsilon v k_j \frac{1}{12 \tau |\varepsilon {\tilde \varepsilon}_0|} \biggr), 
\end{equation}
where $|{\tilde \varepsilon}_0| = |\varepsilon| + 1/(2 \tau)$. As stressed elsewhere \cite{NI2}, the divergent behavior proportional to $|\varepsilon|^{-1}$ in the second term is a consequence of cancellation between the quasiparticle relaxation rate and the impurity-ladder vertex correction and is the origin of the impurity scattering-independent transition temperature in the $s$-wave superconductor. In this sense, this cancellation may be regarded as one analog of the Anderson's theorem \cite{Anderson,Werthamer} in the $s$-wave superconductor. 

In the opposite limit where $\delta \to +\infty$, i.e., the limit of strong anisotropy, the coefficients in $\Lambda_j$ have the following limiting values: 
\begin{eqnarray}
B_0 &\simeq& - \Delta B = - D_0 = \frac{\pi}{16 |\varepsilon {\tilde \varepsilon}_\infty| \tau}, \nonumber \\
\Delta D &\simeq& - D_0 + \frac{\pi}{8 \varepsilon^2 \tau}, \nonumber \\
C_0 &\simeq& \frac{|{\tilde \varepsilon}_\infty|}{|\varepsilon|}, \nonumber \\
C_{21} &\simeq& - \frac{\pi}{32 \varepsilon^2 |{\tilde \varepsilon}_\infty| \tau}, \nonumber \\
C_{1z} &\simeq& - \frac{\pi}{16 {\tilde \varepsilon}_\infty^2 |\varepsilon| \tau} \biggl( 1 + \frac{\pi}{8 |\varepsilon| \tau} \biggr), 
\end{eqnarray} 
and $C_{2z} = - C_{21} - C_{1z}$, where $|{\tilde \varepsilon}_\infty| = |\varepsilon| + \pi/(4 \tau)$. 
Then, $\Lambda_j$ approaches 
\begin{eqnarray}
\Lambda^{(\infty)}_j &=& \delta^{({\rm z})}_{ij} ({\hat p}_j - {\rm i} s_\varepsilon \frac{\pi}{16 |\varepsilon {\tilde \varepsilon}_\infty| \tau} v k_j {\hat p}_\perp^2 ) 
+ {\hat z}_i {\hat z}_j \frac{|{\tilde \varepsilon}_\infty|}{|\varepsilon|} \nonumber \\ 
&\times& \biggl[ {\hat p}_j \biggl( 1 - \frac{\pi}{32 {\tilde \varepsilon}_\infty^2 |\varepsilon|\tau} v^2 k_\perp^2 \biggr) - {\rm i} s_\varepsilon {\hat p}_z^2 \frac{\pi}{8 |\varepsilon {\tilde \varepsilon}_\infty| \tau} v k_j \biggr] \nonumber \\
&-& \delta^{({\rm z})}_{ij} {\bf k}_j \frac{\pi}{16 {\tilde \varepsilon}_\infty^2 |\varepsilon| \tau} \biggl( 1 + \frac{\pi}{8 |\varepsilon| \tau} \biggr) v^2 {\hat p}_z k_z 
\end{eqnarray}
where ${\hat p}_\perp^2 = 1 - {\hat p}_z^2$. We note that the second term of $|{\tilde \varepsilon}_\infty|$ corresponds to the imaginary part of the self energy of the normal Green's function. The $k_\perp^2$ term in eq.(11) suggests the presence of a diffusion pole $(2|\varepsilon| + v^2 \tau k^2/\pi)^{-1}$. 
Consequences of $\Lambda^{(\infty)}_j$ are reflected in each term of the GL free energy which will be given in the next section and Appendices.

\section{Resulting GL free energy}

We will use an appropriate GL free energy to numerically study the vortex solutions stable in anisotropic aerogels. As has been assumed in Ref.\cite{NI1}, the terms arising from spatial variations of the superfluid transition temperature $T_c$ and acting as a pinning potential of a vortex will be neglected in the free energy terms written by the order parameter field $A_{\mu i}$. Further, any term associated with the repulsive channel of the quasiparticle interaction will be neglected in this section. Then, the GL free energy $F_{\rm GL} = F_2 + F_4$ in the presence of the impurity-scattering effect, as usual, consists of the quadratic term 
\begin{eqnarray}
F_{2}/\Omega &=& \sum_{\bf q} \biggl[\frac{N(0)}{3} \biggl({\rm ln}\biggl(\frac{T}{T_{c0}}\biggr) + T \sum_\varepsilon \frac{\pi}{|\varepsilon|} \biggr) \delta_{i,j} \nonumber \\
&-& T \sum_\varepsilon \int_{\bf p} {\hat p}_i \Lambda_j({\hat {\bf p}}, {\bf q}) {\cal G}_\varepsilon({\bf p}+{\bf q}/2) {\cal G}_{-\varepsilon}(-{\bf p}+{\bf q}/2) \biggr]\nonumber \\
&\times& A^*_{\mu i}({\bf q}) A_{\mu j}({\bf q}),
\end{eqnarray}
and the quartic term 
\begin{widetext}
\begin{eqnarray}
F_4/\Omega &=& \frac{1}{2}\left[A^*_{\mu i}A_{\mu j}A^*_{\nu k}A_{\nu l}-A^*_{\mu i}A_{\nu j}A^*_{\mu k}A_{\nu l}+A^*_{\mu i}A_{\nu j}A^*_{\nu k}A_{\mu l}\right] \nonumber \\ 
&\times& T \sum_{\varepsilon} \int_{\mathbf{p}_1} \int_{\mathbf{p}_2} \int_{\mathbf{p}_3} \int_{\mathbf{p}_4} \Lambda_i ({\hat {\bf p}}_1;0) \Lambda_j ({\hat {\bf p}}_2;0) \Lambda_k ({\hat {\bf p}}_3;0) \Lambda_l ({\hat {\bf p}}_4;0) {\cal G}_{\varepsilon}\left({\bf p}_1 \right) {\cal G}_{-\varepsilon}\left(-{\bf p}_1 \right) {\cal G}_{\varepsilon}\left({\bf p}_3 \right) {\cal G}_{-\varepsilon}\left(-{\bf p}_3 \right)\nonumber\\
&\times&\left(\delta_{{\bf p}_1, {\bf p}_2} \delta_{{\bf p}_1, {\bf p}_3} \delta_{{\bf p}_1, {\bf p}_4} +\delta_{{\bf p}_1, {\bf p}_4} \delta_{{\bf p}_2, {\bf p}_3} {\cal G}_{\varepsilon}\left({\bf p}_1 \right) {\cal G}_{\varepsilon}\left({\bf p}_2 \right)\overline{\left|u_{{\bf p}_1-{\bf p}_2}\right|^2}+\delta_{{\bf p}_1, {\bf p}_2}\delta_{{\bf p}_3, {\bf p}_4} {\cal G}_{\varepsilon}\left({\bf p}_1 \right) {\cal G}_{\varepsilon}\left({\bf p}_3 \right)\overline{\left|u_{{\bf p}_1-{\bf p}_3}\right|^2}\right), \nonumber\\
\end{eqnarray}
\end{widetext}
where $\Omega$ is the volume, and, for simplicity, $F_4$ was written here by assuming the order parameter to be spatially 
uniform. 
Up to the lowest order in the spatial gradient, one can separate $F_2 + F_4$ into the bulk energy contribution $F_{\rm {bulk}} = \int_{\bf r} f_{\rm {bulk}}$, where 
\begin{widetext}
\begin{eqnarray}
f_{\rm{bulk}}&=&(\alpha+(\alpha_z-\alpha)\delta_{iz})A_{\mu i}A^*_{\mu i}+\beta_1^{(0)}|A_{\mu i}A_{\mu i}|^2+\beta_2^{(0)}(A_{\mu i}A_{\mu i}^*)^2+\beta_3^{(0)} A_{\mu i}^*A_{\nu i}^*A_{\mu j}A_{\nu j} \nonumber\\
&&+\beta_4^{(0)} A_{\mu i}^*A_{\nu i}A_{\nu j}^*A_{\mu j}+\beta_5^{(0)} A_{\mu i}^*A_{\nu i}A_{\nu j}A_{\mu j}^*+\beta_z |A_{\mu z}A_{\mu z}^*|^2\nonumber\\
&&+[\beta_{1}^{(1)}A_{\mu i}A_{\mu i}A_{\mu z}^*A_{\mu z}^*+\beta_{2}^{(1)} A_{\mu i}A_{\mu i}^*A_{\mu z}A_{\mu z}^*+\beta_{3}^{(1)} A_{\mu i}^*A_{\nu i}^*A_{\mu z}A_{\nu z} \nonumber\\
&&+\beta_{4}^{(1)} A_{\mu i}^*A_{\nu i}A_{\nu z}^*A_{\mu z}+\beta_{5}^{(1)} A_{\mu i}^*A_{\nu i}A_{\nu z}A_{\mu z}^* +c.c.], 
\end{eqnarray}
\end{widetext}
and the gradient terms. In the case of weak anisotropy where $\delta \ll 1$, the $\beta_{n}^{(1)}$ ($n=1, \cdot \cdot \cdot , 5$) terms appear in O($\delta$), while $\beta_z$ term first appears in O($\delta^2$). 

Among the gradient terms, the free energy density corresponding to the contributions from $F_2$ consist of the following seven terms: 
\begin{eqnarray}
f_{\rm{grad}} &=& 2K_1\partial_i A_{\mu i}\partial_j A_{\mu j}^* +K_2\partial_i A_{\mu j}\partial_i A_{\mu j}^*
+K_3\partial_z A_{\mu i}\partial_z A_{\mu i}^* \nonumber\\
&+&  K_4\partial_i A_{\mu z}\partial_i A_{\mu z}^*
+K_5(\partial_i A_{\mu i}\partial_z A_{\mu z}^* + \rm{c.c.}) \nonumber \\
&+& K_6 \partial_z A_{\mu z}\partial_z A_{\mu z}^*. 
\end{eqnarray}

General expressions on the coeffients in the GL free energy are involved and will be presented in Appendix. Here, their limiting behaviors in the limits of weak anisotropy, $\delta \to 0$, and of strong anisotropy, $\delta \to \infty$ will be explained together with their implication. In $\delta \to 0$ limit, up to O($\delta$), $\beta_z$ vanishes, and the remaining GL coefficients of $f_{\rm bulk}$ reduce to those given in Ref.\cite{AI06}. In the isotropic ($\delta \to 0$) limit, the four coefficients of $f_{\rm {grad}}$, $K_j$ ($j=3, \cdot\cdot\cdot, 6$), vanish, while $K_1$ and $K_2$ coincide with those given in Ref.\cite{NI2}. Among them, $K_1$ logarithmically diverges upon cooling, reflecting the cancellation between the relaxation rate and the pairing vertex mentioned in sec.II.  

In contrast, in the $\delta \to \infty$ limit, the coefficients in $f_{\rm bulk}$ have the following limiting values : 
\begin{eqnarray}
\alpha &\simeq& \frac{1}{3} N(0) \biggl[ {\rm ln}\biggl(\frac{T}{T_{c0}} \biggr) + \psi\biggl(\frac{1}{2}+\frac{1}{8 \tau T} \biggr) - \psi\biggl(\frac{1}{2} \biggr) \biggr], \nonumber \\
\alpha_z &\simeq& \frac{1}{3} N(0) {\rm ln}\biggl(\frac{T}{T_{c0}} \biggr), \nonumber \\
\end{eqnarray}

\begin{eqnarray}
\beta_3^{(0)} &=& -2 \beta_1^{(0)} \simeq \frac{\pi T}{15} N(0) \sum_{\varepsilon > 0} \frac{1}{|{\tilde \varepsilon}_\infty|^3}, \nonumber \\
\beta_2^{(0)} &=& \beta_4^{(0)} = - \beta_5^{(0)} \simeq \beta_3^{(0)} - \frac{\pi^2 T}{60 \tau} N(0) \sum_{\varepsilon > 0} \frac{1}{{\tilde \varepsilon}_\infty^4}, \nonumber \\
\beta_3^{(1)} &=& -2 \beta_1^{(1)} \simeq - \beta_3^{(0)} + \frac{\pi T}{15} N(0) \sum_{\varepsilon > 0} \frac{1}{\varepsilon^2 |{\tilde \varepsilon}_\infty|}, \nonumber \\
\beta_2^{(1)} &=& \beta_4^{(1)} = - \beta_5^{(1)} \simeq \beta_3^{(1)} + \frac{\pi^2 T}{60 \tau} N(0) \sum_{\varepsilon > 0} \frac{1}{{\tilde \varepsilon}_\infty^4} \biggl( 1 - \frac{{\tilde \varepsilon}_\infty^2}{2 \varepsilon^2} \biggr), \nonumber \\
\beta_{12345}^{(0)} &+& 2 \beta_{12345}^{(1)} + \beta_z \simeq \frac{\pi T}{10} N(0) \sum_{\varepsilon > 0} \frac{1}{|\varepsilon|^3}, 
\end{eqnarray}
where $\beta_{12345}=\beta_1+\beta_2+\beta_3+\beta_4+\beta_5$. It can be verified that $\beta^{(1)}_n$ and $\beta_z$ vanish in the true clean limit 
where $\tau^{-1}=0$. 

The fact that $\alpha_z$ approaches its result in the impurity-free bulk liquid is a consequence of the specular scattering along the polar axis (see eq.(5)) and implies that the superfluid transition temperature $T_c$ between the normal phase and the polar pairing state is not affected by the impurity scattering in the limit of strong anisotropy. Thus, this $\alpha_z$-expression can be regarded as another analog of the Anderson's theorem \cite{Anderson} in the $s$-wave superconductor \cite{Fomin18,Eltsov19,IT19}. Consistently with the behavior of $\alpha_z$, the last line of eq.(18) which is the coefficient of the quartic bulk term associated with the description of the polar phase also becomes independent of $\tau$. Therefore, the squared amplitude of the order parameter $|\Delta_{\rm polar}|^2 = A^*_{\mu z} A_{\mu z}$ which becomes $- \alpha_z/(\beta_{12345}^{(0)} + 2 \beta_{12345}^{(1)} + \beta_z)$ within the present GL treatment is also independent of $\tau$ \cite{Eltsov19}. 
Such an impurity-free nature at the polar to normal transition does not hold at the transition temperature to another pairing state at lower temperatures \cite{Fomin18,IT19}. 

As stressed elsewhere \cite{IT19}, the above-mentioned impurity-free thermodynamic behavior of the polar pairing state at finite temperatures is approximatedly seen if $\delta \geq 5$. So, even in aerogels to be modelled by a finite $\delta$, the model in the limit of strong anisotropy can be conveniently used for theoretical descriptions.  

Due to the nonvanishing $\delta$, as seen in eq.(16), the quadratic gradient energy $f_{\rm grad}$ consists of the six invariants, and the corresponding six coefficients remain nonvanishing even in the limit of strong anisoropy ($\delta \to \infty$). In low $T$ limit,  all coefficients remain nonvanishing, and their leading terms in low $T$ limit become 
\begin{eqnarray}
K_1 &\simeq& \frac{\pi^2 T v^2}{120} N(0) \sum_{\varepsilon > 0} \frac{1}{|\varepsilon| \tau |{\tilde \varepsilon}_\infty|^3}, \nonumber \\
K_4 &\simeq& \frac{\pi^2 T v^2}{48} N(0) \sum_{\varepsilon > 0} \frac{1}{|\varepsilon|^2 \tau |{\tilde \varepsilon}_\infty|^2}, \nonumber \\
K_5 &\simeq& \frac{\pi^2 T v^2}{480} N(0) \sum_{\varepsilon > 0} \frac{1}{|\varepsilon|^2 \tau |{\tilde \varepsilon}_\infty|^2} \biggl( 1 + \frac{5}{4} \frac{\pi}{|{\tilde \varepsilon}_\infty| \tau} \biggr). \nonumber \\
\end{eqnarray}
Further, the coefficient of $\partial_z A_{\mu z}^* \partial_z A_{\mu z}$ which arises from the sum of the $K_4$, $K_5$, and $K_6$ terms approaches 
\begin{equation}
\frac{\pi^2 T v^2}{80} N(0) \sum_{\varepsilon > 0} \frac{1}{\varepsilon^2 {\tilde \varepsilon}_\infty^2 \tau}.
\end{equation}
The divergent behaviors $\sim \varepsilon^{-2}$ in low $T$ limit of eq.(20) corelates with the $\tau$-independent $\sim - |\varepsilon|^{-1}$ behavior leading to the $|{\rm ln}T|$ contribution in $\alpha_z$. In contrast, $K_2$ and $K_3$ reduce to finite values in low $T$ limit.

\section{Description of HQV pair in PdB phase in London limit}

To correctly understand the order parameter structures of a HQV pair obtained numerically, it is useful to have an intuitive image of a HQV pair by describing it in the London limit where the order parameter $A_{\mu j}$ is described in terms of the angle variables while keeping the overall amplitude fixed. Hereafter,  we focus on the HQV lines extended along the polar axis ${\hat z}$. 

First, let us review how to describe a single HQV \cite{Volovik90}. By expressing a relative rotation around the $x$-axis between the orbital and spin frames in terms of the rotaion matrix $(R_x(\theta))_{\mu \nu} = {\hat x}_\mu {\hat x}_\nu + \delta^{({\rm x})}_{\mu \nu} {\rm cos}\theta - \varepsilon_{x \mu \nu} {\rm sin}\theta$, the order parameter in the PdB phase in an environment with a uniaxially stretched anisotropy is expressed following the notation in Ref.\cite{Yang} as 
\begin{eqnarray}
A_{\mu j} &=& |\Delta| e^{{\rm i}\Phi} (R_x(\theta))_{\mu \nu} \biggl(\frac{c}{\sqrt{2}} \delta^{({\rm z})}_{\nu j} + \sqrt{1 -c^2} {\hat z}_\nu {\hat z}_j \biggr) \nonumber \\
&=& \frac{|\Delta|}{\sqrt{2}} e^{{\rm i}\Phi} 
\begin{pmatrix}
c & 0 & 0  \\
0 & c \, {\rm cos}\theta & -\sqrt{2(1 - c^2)} \, {\rm sin}\theta \\
0 & c \, {\rm sin}\theta & \sqrt{2(1 - c^2)} \, {\rm cos}\theta 
\end{pmatrix}, 
\end{eqnarray}
where $c$ ($0 \leq c \leq \sqrt{2/3}$) is the parameter playing the role of the order parameter of the PdB phase, $\delta^{({\rm x})}_{\mu \nu} =\delta_{\mu \nu} - {\hat x}_\mu {\hat x}_\nu$, and the over all phase $\Phi$ was introduced. A single HQV localized at the origin is expressed by choosing $\Phi=\theta=\phi/2$. Then, the corresponding order parameter becomes 
\begin{equation}
A_{\mu j} = \frac{|\Delta|}{\sqrt{2}} \biggl[ c e^{{\rm i}\phi/2} {\hat x}_\mu {\hat x}_j + \sqrt{1 - \frac{c^2}{2}} \biggl( e^{{\rm i}\phi} {\hat e}_{- \mu} {\hat e}'_{+ j} + {\hat e}_{+ \mu} {\hat e}'_{- j} \biggr) \biggr], 
\end{equation}
where the unit vectors ${\hat e}_{\pm \mu} = ({\hat y} \pm {\rm i} {\hat z})_\mu/\sqrt{2}$, and ${\hat e}'_{\pm j} = (c {\hat y} \pm {\rm i} \sqrt{2(1-c^2)} {\hat z})_j/\sqrt{2-c^2}$ were introduced. The fact that only $A_{xx}$ does not become a single-valued component upon circling the vortex center implies that, as sketched in Fig.2(a), $A_{xx}$ inevitably vanishes on a wall corresponding to a branch cut with a fixed $\phi$-value. In other words, a polar-distorted {\it planar} phase is realized on the wall \cite{Volovik90,Volovik19}. 
\begin{figure}[t]
{
\includegraphics[scale = 1.4]{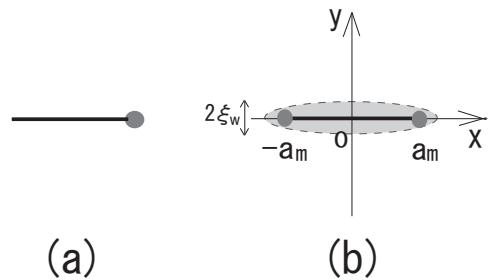}
}
\caption{(a) A single HQV (solid dot) in a B phase described in the $x$-$y$ plane. To make the order parameter $A_{\mu,j}$ single-valued, its one component must vanish on the string (solid line). (b) A pair of HQVs in a B phase is accompanied by the string on which $A_{\mu,j}$ becomes the two-dimensional planar state. This planar string has the length $\simeq 2 a_m$ and a width $2 \xi_w$ if this HQV pair is well defined. In the double-core vortex in the bulk liquid, the string shrinks, and the planar state appears only at the center of the vortex, i.e, the origin (see sec.V). 
}
\label{s:fig:HQV}
\end{figure}
The above expression of the order parameter in the London limit is easily generalized to the case with a HQV pair. Since we should consider a HQV pair to be compared with the ordinary phase vortex with an integer winding number of the phase, the angle variables $\Phi$ and $\theta$ will be chosen in the manner 
\begin{equation}
\Phi = \frac{\phi_++\phi_-}{2}, \,\,\,\,\,\ \theta = \frac{\phi_+-\phi_-}{2}
\end{equation}
where $\phi_\pm = {\rm tan}^{-1}[y/(x \mp a)]$. Then, eq.(22) is replaced by eq.(21) with eq.(23), i.e.,  
\begin{eqnarray}
A_{\mu j} &=& \frac{|\Delta|}{\sqrt{2}} \biggl[ c e^{{\rm i}(\phi_++\phi_-)/2} {\hat x}_\mu {\hat x}_j + \sqrt{1 - \frac{c^2}{2}} \biggl( e^{{\rm i}\phi_+} {\hat e}_{- \mu} {\hat e}'_{+ j} \nonumber \\
&+& e^{{\rm i}\phi_-} {\hat e}_{+ \mu} {\hat e}'_{- j} \biggr) \biggr]. 
\label{matrixLondon}
\end{eqnarray}
In this case sketched in Fig.2(b), the expression of $A_{xx}$ impies that the order parameter in $|x| > a$ is continuous through the $x$-axis, while the wall is necessary in $|x| < a$. 
In the polar limit where $c \to 0$, this expression reduces to the order parameter of the polar phase with the $d$-vector $d_\mu = {\hat z}_\mu {\rm cos}\theta - {\hat y}_\mu {\rm sin}\theta$. 

Next, the dependence of the HQV pair's energy on the HQV pair size $2a$ will be considered using the gradient energy terms. The $a$-dependent contribution of the vortex energy will be denoted as $\Delta F_{\rm L}(a) = F(a) - F(\xi_c)$, where $\xi_c$ is a cut off length corresponding to the core size of a HQV over which the London limit may be used. 
Since only the vortex lines straight along the $z$-axis are considered, 
any gradient terms including the gradient $\partial_z$ are neglected. If our attention is paid only to such terms in the quadratic gradient energy of eq.(16), the corresponding $\Delta F_{\rm L}(a)$ 
becomes 
\begin{equation}
\Delta F_{\rm L}^{(2)}(a) = - \frac{\pi}{2} c^2 |\Delta|^2 (K_1+K_2) {\rm ln}\biggl(\frac{2a}{\xi_c} \biggr) 
\end{equation}
in clean limit where $\tau^{-1}=0$, where $K_1+K_2 \simeq 7 \zeta(3) \xi_0^2 N(0)T_{c0}^2/(30 T^2)$, and $\xi_0 = v/(2 \pi T_{c0})$. 
The corresponding results in the two limits of eq.(25) are already known: Such an energy gain of the double-core vortex relative to the so-called $o$-vortex \cite{Fujita} in the bulk B phase is given by eq.(25) with $c^2=2/3$ \cite{Volovik90,NI2}. Further, the factor $c^2$ in eq.(25) is consistent with the vanishing $\Delta F_{\rm L}^{(2)}(a)$ in the opposite polar limit where $c=0$ \cite{NI1}. It can be checked that the nonvanishing eq.(25) proportional to $c^2$ follows only from spatial variations of $A_{xx}$ which is negligible in the polar phase. As shown in Ref.\cite{NI1}, the negative $\Delta F_{\rm L}(a)$ in the polar limit occurs only from the gradient term expressing the Fermi-liquid (FL) or spin-fluctuation correction, and the corresponding contribution to $\Delta F_{\rm L}(a)$ is given, in the FL model, by \cite{NI1}
\begin{eqnarray}
\Delta F_{\rm L}^{(4)}(a) &\simeq& \frac{c^{-2}}{30} \Gamma_1^s |\psi^{(2)}(1/2)| \biggl(\frac{T_{c0} |\Delta|}{\pi T^2} \biggr)^2 \Delta F_{\rm L}^{(2)}(a) 
\nonumber \\
&\simeq& - 0.1 \pi \Gamma_1^s \biggl(\frac{T_{c0} |\Delta|}{\pi T^2} \biggr)^2 \xi_0^2 N(0) |\Delta|^2 {\rm ln}\biggl(\frac{2a}{\xi_c} \biggr), 
\end{eqnarray}
where the next order terms of O($c^2$) were neglected. Here, $\psi^{(2)}(1/2) = - 14 \zeta(3)$, and $\Gamma_1^s = F_1^s/(1 + F_1^s/3)$ is the pressure-dependent constant of order unity with a Landau parameter $F_1^s (> 0)$. Thus, the energy gain corresponding to an attractive force in a HQV pair is dominated by the FL correction term rather than the ordinary weak-coupling terms in the strongly anisotropic PdB phase with a low enough $|c|$-value. 

In the present PdB phase, we also have an energy cost due to the planar wall. This contribution due to the nonzero $A_{xx}$ is estimated like 
\begin{equation}
\Delta F_w(a) \simeq 0.1 N(0) \xi_0^2 |\Delta \, c|^2 a/\xi(T),  
\end{equation}
where $\xi(T) = \xi_0(N(0)/|\alpha|)^{1/2}$. 
The coefficient $\Delta F_w(a)/a$ measures the line tension of the wall per unit length. By optimizing the sum $\Delta F_{\rm L}^{(2)}+\Delta F_{\rm L}^{(4)}+\Delta F_w$ w.r.t. $a$, the pair size to be realized is given by 
\begin{equation}
2 a_{m} \simeq c^{-2} \frac{2 \Gamma_1^s}{\pi} \biggl(\frac{|\Delta|}{T} \biggr)^2 \xi(T). 
\end{equation}
In this way, It is expected in the London limit that the size of a HQV pair, $a_m$, i.e., the longer radius of the elliptical core of the double-core vortex, is a microscopic scale in the bulk B phase, while, in a strongly anisotropic PdB phase close to $T_{\rm PB}$ where $|c| \ll 1$, the pair size may become a macroscopic one. This result will be used to explain the content of our numerical results in the next section. 

If, as in the conventional GL approach \cite{Yang,Fujita,Th87,SVreview,VW,Th98}, the FL corrected gradient term is neglected, the HQV pair size would remain microscopic so that the presence \cite{Eltsov} of a macroscopic HQV pair in the PdB phase could not be explained. The appearance of a macroscopic HQV pair in the PdB phase is a combined effect of a strong anisotropy {\it and} the FL correction to the free energy. 

\vspace{5mm}
\section{Numerical analysis and results} 

In our numerical study, the GL model we use consists of the three contributions to the free energy density, $f_{\rm bulk}$, $f_{\rm grad}$, defined in sec.III, and the additional O($|\Delta|^4$) contributions $f_{\rm grad}^{(4)}$. Before proceeding to discussing about our numerical results, comments on the O($|\Delta|^4$) gradient terms \cite{NI1} have to be given.

\begin{figure}[b]
{
\includegraphics[scale = 1.4]{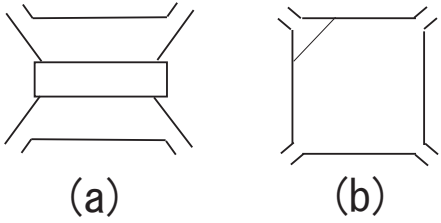}
}
\caption{(a) Diagram giving the FL-correction to the gradient energy. The rectangle denotes the vertex part representing the renormalized interaction between the quasiparticles. (b) One example of a vertex correction to the weak coupling quartic order term (Gor'kov box). 
}
\label{s:fig:4thDiagram}
\end{figure}
Among the two terms composing $f_{\rm grad}^{(4)}$, the stability of HQVs in the polar and A phases is determined by the contribution arising from the repulsive channel of the interaction between the quasiparticles to $f_{\rm grad}^{(4)}$. Hereafter, to simplify our description, we will use the Fermi liquid (FL) model of such an interaction contribution to $f_{\rm grad}^{(4)}$ for our numerical study. The corresponding gradient energy contribution $f_{\rm FL}^{(4)}$, described in Fig.3(a), has been derived in Ref.\cite{NI1} by neglecting the anisotropy effects and, in the limit of strong anisotropy, takes the form  
\begin{widetext}
\begin{eqnarray}
f_{FL}^{(4)} \!\!\! &=& \frac{N(0)}{450} \Gamma_1^s (\pi v)^2 \biggl( T \sum_{\varepsilon > 0} \frac{1}{{\tilde \varepsilon}_\infty^3} \biggr)^2  \biggl[ \! (\nabla\cdot A_\mu)(\nabla \cdot A_\lambda^*)A_{\mu i}^*A_{\lambda i} 
+ (\nabla A_{\mu i}) \cdot (\nabla A_{\lambda j}^*)A_{\mu i}^*A_{\lambda j} 
 + (A_\lambda \cdot \nabla)A_{\lambda i}^*(A_\mu^* \cdot \nabla)A_{\mu i} \nonumber \\
&-& \biggl( (\nabla A_{\mu i}^*) \cdot (\nabla A_{\lambda j}^*)A_{\mu i}A_{\lambda j} + (A_{\mu} \cdot \nabla)A_{\mu i}^*(A_{\lambda} \cdot \nabla)A_{\lambda i}^* + (\nabla \cdot A_\mu^*)(\nabla \cdot A_\lambda^*)A_{\mu i} A_{\lambda i} \biggr) \nonumber \\
&+& 2 \biggl[ (\nabla \cdot A_\mu)(A_{\mu i}^*(A_\lambda \cdot \nabla)A_{\lambda i}^* + A_{\lambda i}(A_{\mu}^* \cdot \nabla)A_{\lambda i}^*) + 
A_{\mu i}^*((A_\lambda \cdot \nabla)A_\lambda^* \cdot \nabla)A_{\mu i} \nonumber \\
&-& 
\biggl((\nabla \cdot A_\mu^*)(A_{\lambda i}(A_\mu\cdot\nabla)A_{\lambda i}^* 
+ A_{\mu i}(A_\lambda \cdot \nabla)A_{\lambda i}^* ) + A_{\lambda i}((A_\mu \cdot \nabla)A_\mu^* \cdot \nabla)A_{\lambda i}^* \biggr) \biggr] + {\rm c.c.} 
\biggr],  
\end{eqnarray}
\end{widetext}
where the spin-antisymmetric Landau parameter $\Gamma_1^a$ was assumed to be negligibly small \cite{NI1}. The corresponding expression in the isotropic case where $\delta=0$ is given by replacing ${\tilde \epsilon}_\infty$ in eq.(29) by ${\tilde \epsilon}_0$. Although eq.(29) does not include the anisotropy parameter $\delta$ explicitly, the anisotropy-induced vertex correction with $C_0 - 1$ as a coefficient is, as is explained in Appendix, safely negligible even in the limit of strong anisotropy. 

Another contribution to $f_{\rm grad}^{(4)}$ arises from the ordinary weak-coupling O($|\Delta|^4$) term, the so-called "Gor'kov box", unaccompanied by a repulsive interaction between quasiparticles (see Fig.3(b)). This contribution includes all the terms including those expressed by $C_{21}$, $B_0$, and $\Delta B$, in the vertex correction $\Lambda_j$. As is explained in relation to Fig.7(a), however, these vertex corrections are also safely negligible. This has been concluded through the full numerical results, although it is already known \cite{NI1} that these anisotropy-induced terms unaffect the resulting size of the HQV pair irrespective of the anisotropy value. Therefore, regarding $f_{\rm grad}^{(4)}$ to be added to $f_{\rm bulk}$ and $f_{\rm grad}$, its expression in the isotropic case, i.e., eq.(52) in Ref.\cite{NI1} has been used to obtain numerical results even in the case with a strong enough anisotropy. 

To numerically examine how the double-core vortex becomes stable as a HQV pair, we follow the previous work on the double-core vortex in the B phase in the isotropic aerogel \cite{NI2}: First, eq.(24) is used as the initial condition for searching a half-core pair with the lowest energy at fixed values of the temperature and pressure. This London solution, eq.(24), has a fixed size $2a$ of the HQV pair as a parameter. Alternatively, the texture of the order parameter at the outer boundary is initially set by a fixed $a$-value. The variational equations of the GL free energy explained above are solved to obtain the solution minimizing the energy for each $a$-value according to the direct two-dimensional method \cite{Th87}, i.e., by assuming the vortices to be straight line objects extending along the $z$-axis. 
During this procedure, we have checked that the size of the half-core pair of the {\it resulting} double-core vortex solution almost coincides with the $2a$ value at the initial condition. Thus, in examining the dependence of the vortex energy\begin{figure}[htb]
{\includegraphics[scale=0.35]{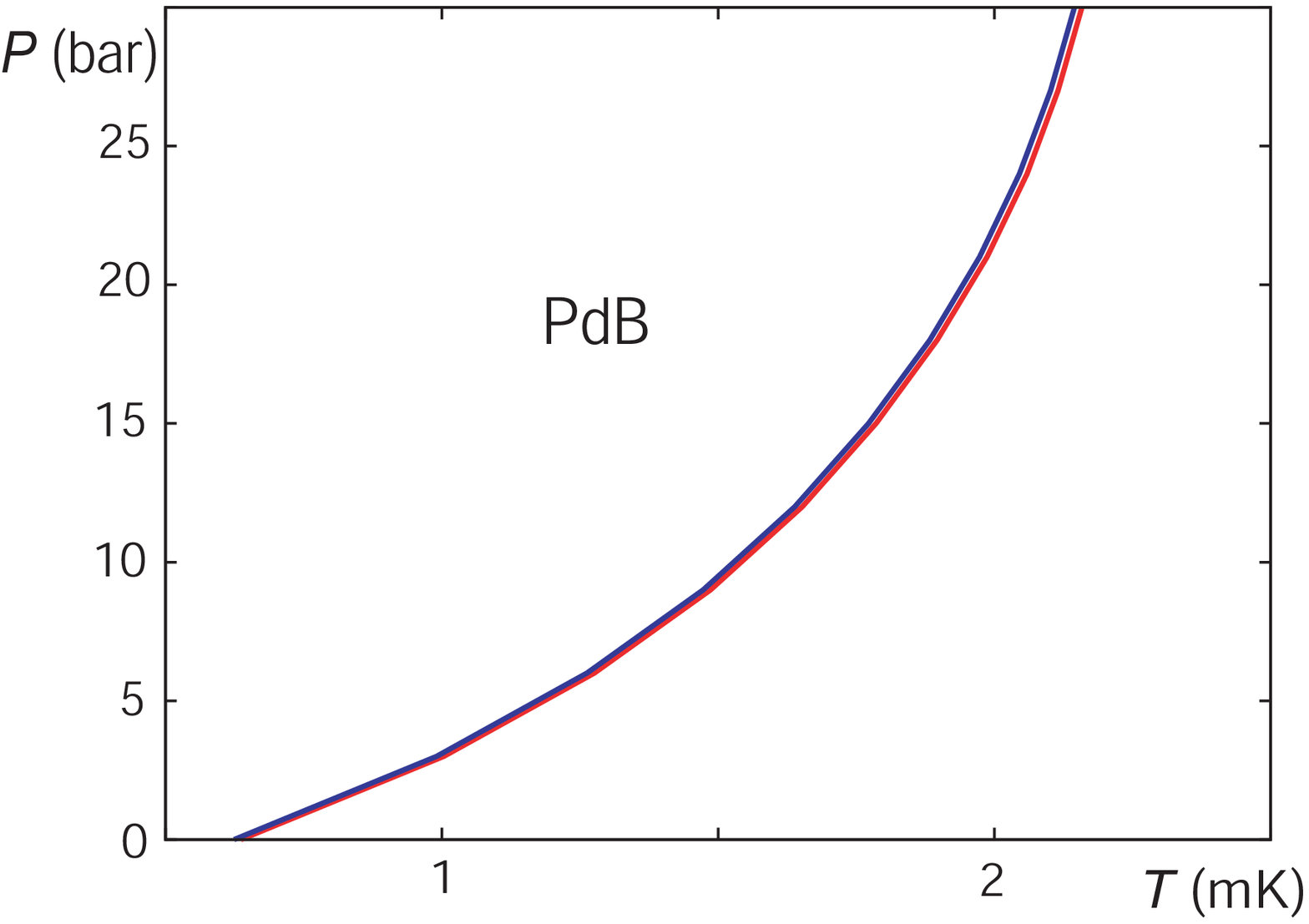}}
{\includegraphics[scale=0.35]{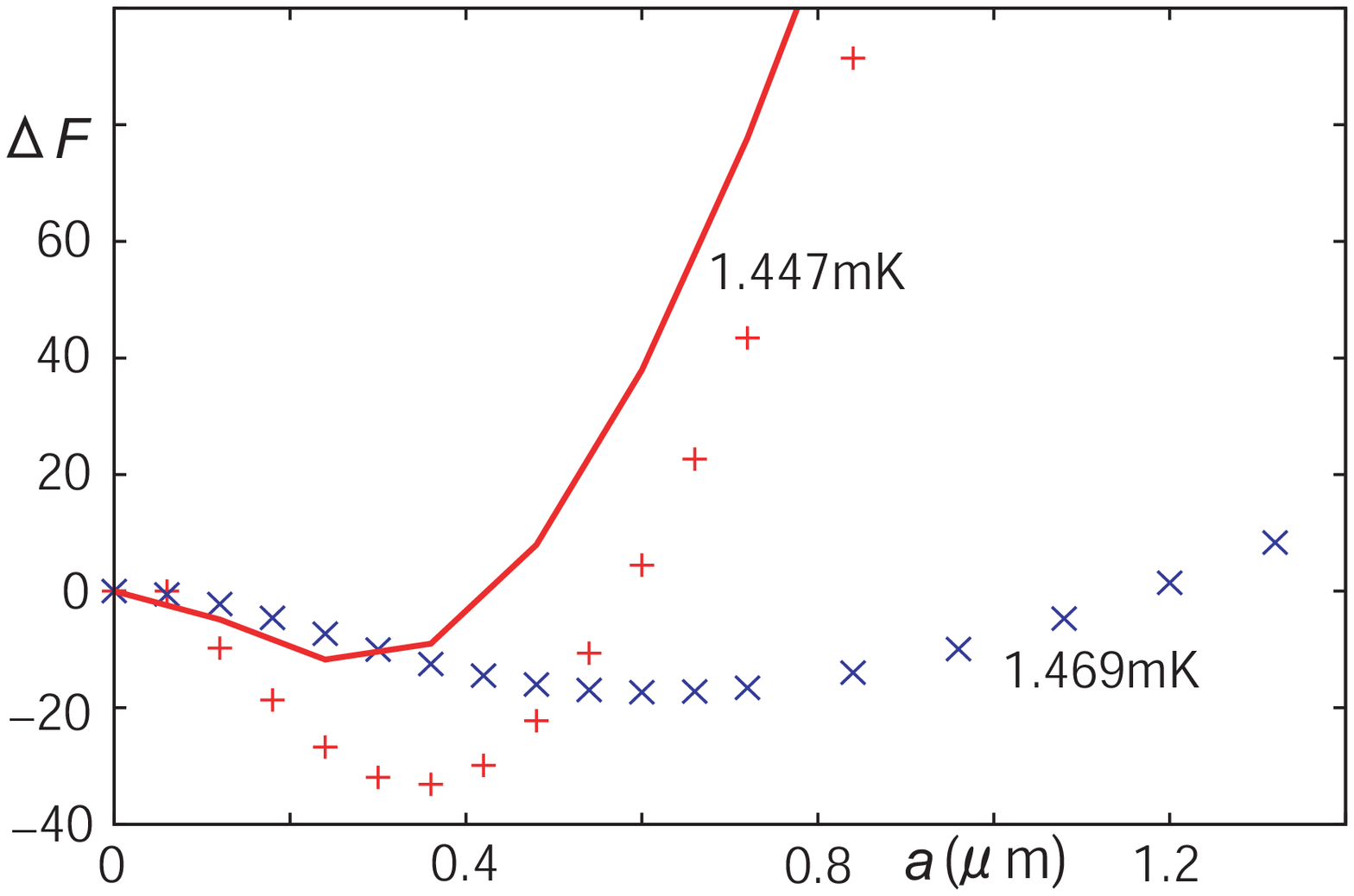}}
\caption{Numerical results for $\delta=0.05$. (a) $P$ v.s. $T$ phase diagram. The red solid curve is the superfluid transition curve $T_c(P)$ between the polar and normal phases, while the blue one is $T_{\rm PB}(P)$ and is quite close to $T_c(P)$. (b) The $a$ v.s. $\Delta F(a)$ curves at $P=9$(bar) and at $T=1.447$ (red plus symbols) and $1.469$ (blue cross symbols) (mK). The $a$-value, $a_m$, minimizing $\Delta F(a)$ in each case is given in Table I to be given later. For comparison, the corresponding curve at 1.447(mK) in the case with no FL correction is shown by the solid red curve which indicate $a_m=0.24$($\mu$m).
}
\label{s:fig:0.05}
\end{figure}
\begin{equation}
\Delta F(a) = F(a) - F(0)
\end{equation}
on the $a$-value introduced as the intial condition below, this $a$ can be identified with the half of the resulting size of the half-core pair. Here, $\Delta F(a)$ corresponds to $\Delta F_L(a)$ introduced in the London limit. In the language of the vortices in the bulk B phase, the $F(0)$ corresponds to the free energy of the so-called $o$-vortex \cite{Fujita}. 

In our computaions studying the vortices extending along the $z$-axis, the system size in the $x$ ($y$) direction was fixed to $24$ ($1.2$) ($\mu$m) in the layout sketched in Fig.2(b). The pressure dependence of the system is incorporated through that of the bulk transition temperature $T_{c0}$ and the Fermi velocity $v$ \cite{VW}. The dimensionless strength of the impurity scattering is $(\tau T_{c0})^{-1}$ which is enhanced with decreasing the pressure reflecting the pressure dependence of $T_{c0}$ \cite{NI2,Th98}. 

\begin{figure}[htb]
{\includegraphics[scale=0.35]{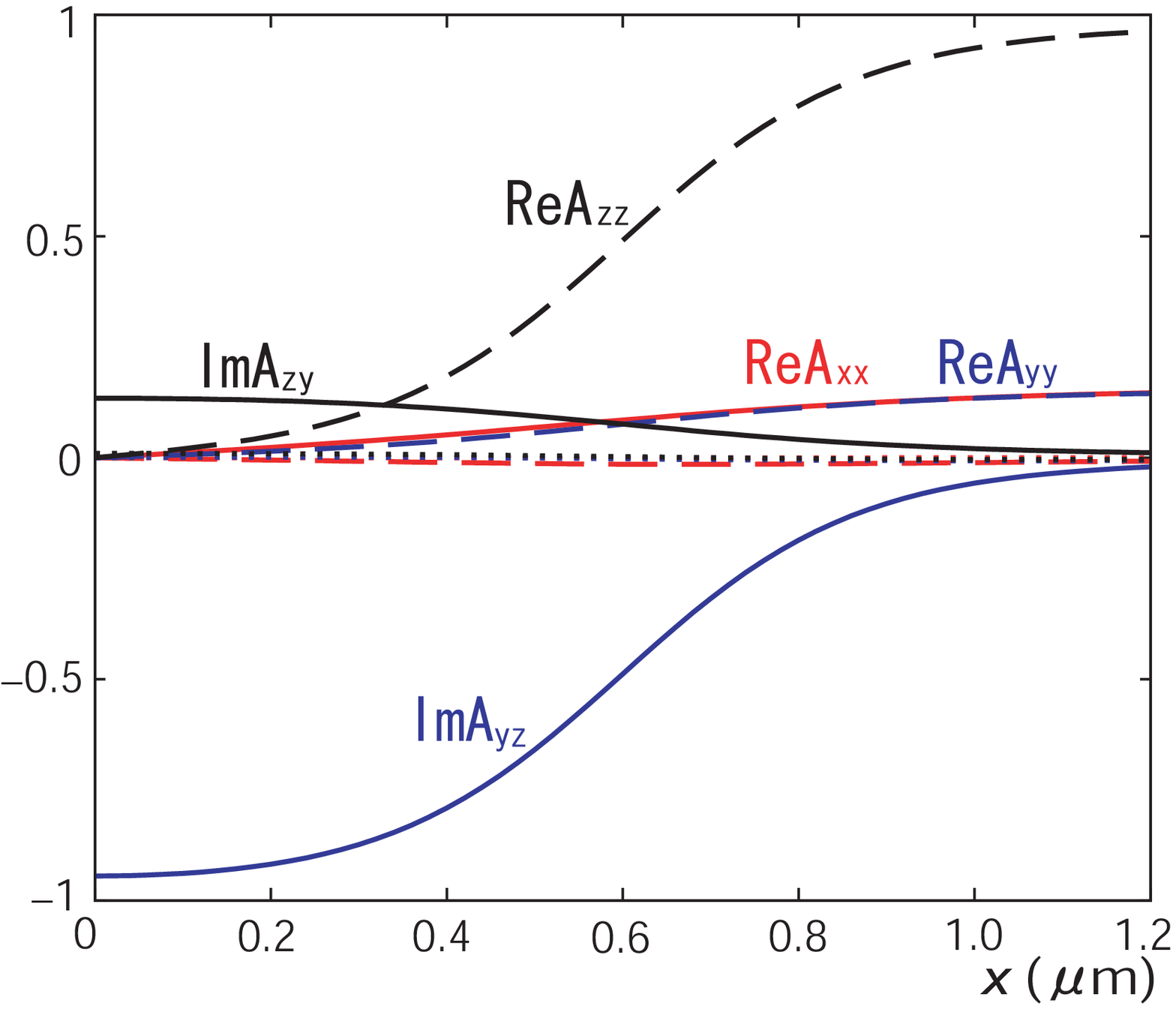}}
{\includegraphics[scale=0.35]{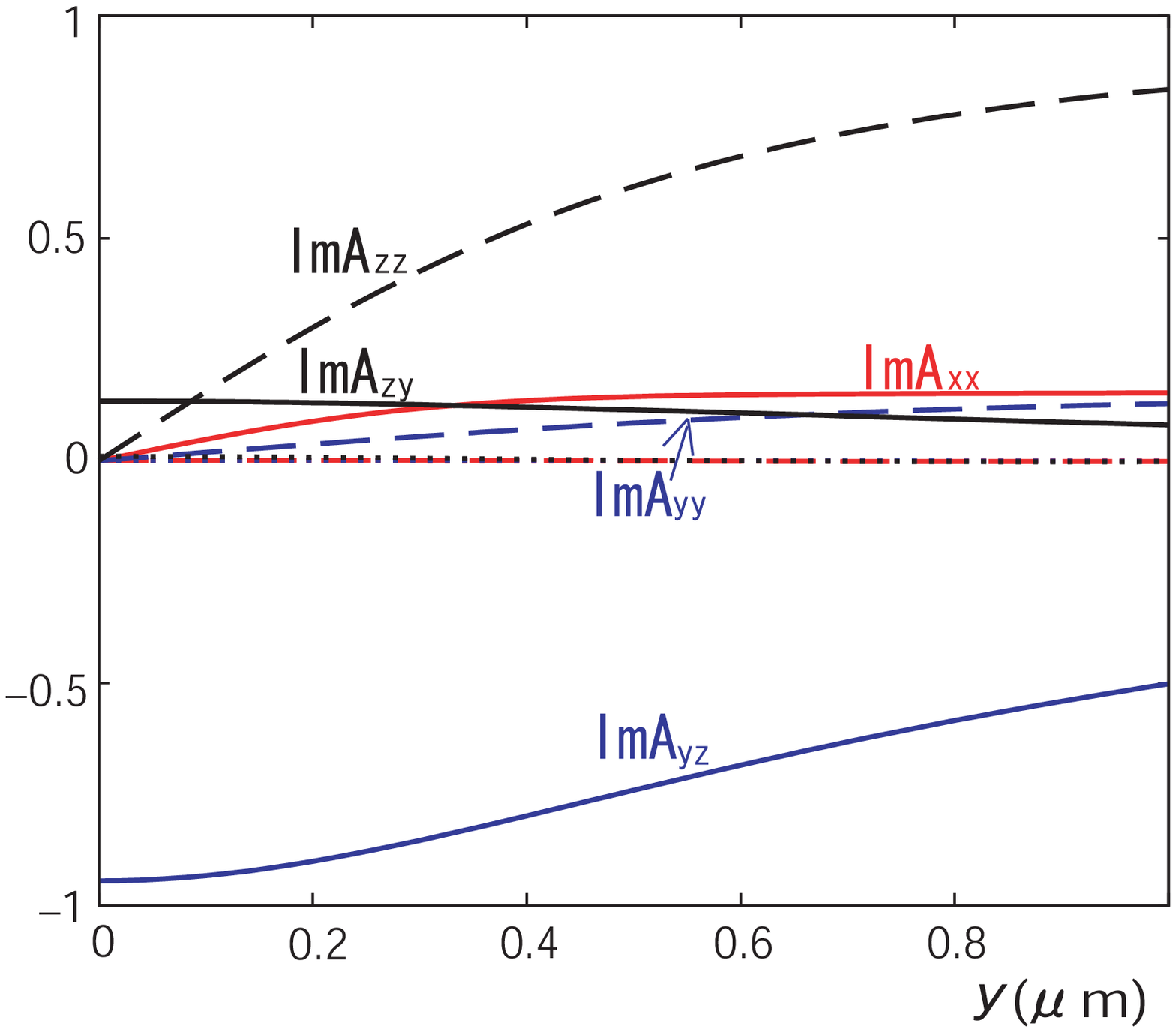}}
\caption{Spatial variations of $A_{\mu j}$ components for $\delta=0.05$ on sweeping (a) along the $x$-axis and at $y=0$ at $a_m \simeq 0.6$($\mu$m) when $T=1.469$(mK) and $P=9$(bar) and (b) along the $y$-axis and at $x=0$. Here, the vortex center is at ($0$, $0$). The $A_{\mu j}$ components other than the nonvanishing five components in the London representation, eq.(21), are expressed by the dotted curves and a red dashed curve.  
}
\label{s:fig:0.05}
\end{figure}

Throughout the present study, the dipole energy is not taken into account. The neglect of the dipole energy is justified in the case of weaker anisotropy where the resulting size of the half-core pair is much smaller than the dipole length $\xi_{\rm D} \sim 10$($\mu$m). In contrast, a HQV pair resulting from a strong enough anisotropy may have a size of the order of $\xi_{\rm D}$ over which the dipole energy affects spatial patterns of the $\theta$-variable, defined in eq.(21), in the PdB phase \cite{Eltsov}. However, one will see below in this section that the London limit becomes a better description as a HQV pair typically grows accompanying the increase of the anisotropy. Then, the dipole energy has only to be taken into account in a decription starting from 
the London limit \cite{Eltsov}. 

Hereafter,as our numerical results at some $\delta$-values, we will present $\Delta F(a)$ data and $x$ and $y$ dependences of each component of the order parameter $A_{\mu,j}$ of the vortex solution minimizing $\Delta F(a)$. The $\tau^{-1}$-value will be fixed to $0.13$(mK) hereafter. First, the $\delta = 0.05$ case is discussed as a typical example of superfluid $^3$He in a {\it weakly anisotropic} aerogel. Figure 4 (a) and (b) express the corresponding phase diagram and the $a$ v.s. $\Delta F(a)$ curves at $T=1.469$ (mK) close to $T_{\rm PB}$ and $1.447$(mK) at a fixed pressure $P=9$ (bar), while Fig.5(a) and (b) present spatial variations of each component of $A_{\mu,j}$ on sweeping along the $x$ and $y$ axis, respectively. Here, the HQV pair is always assumed to be initially set as in Fig.2(b), and the origin is the center of the HQV pair. Further, by symmetry, just the region in $x \geq 0$ and $y \geq 0$ is shown. 
\begin{figure}[t]
{\includegraphics[scale=0.35]{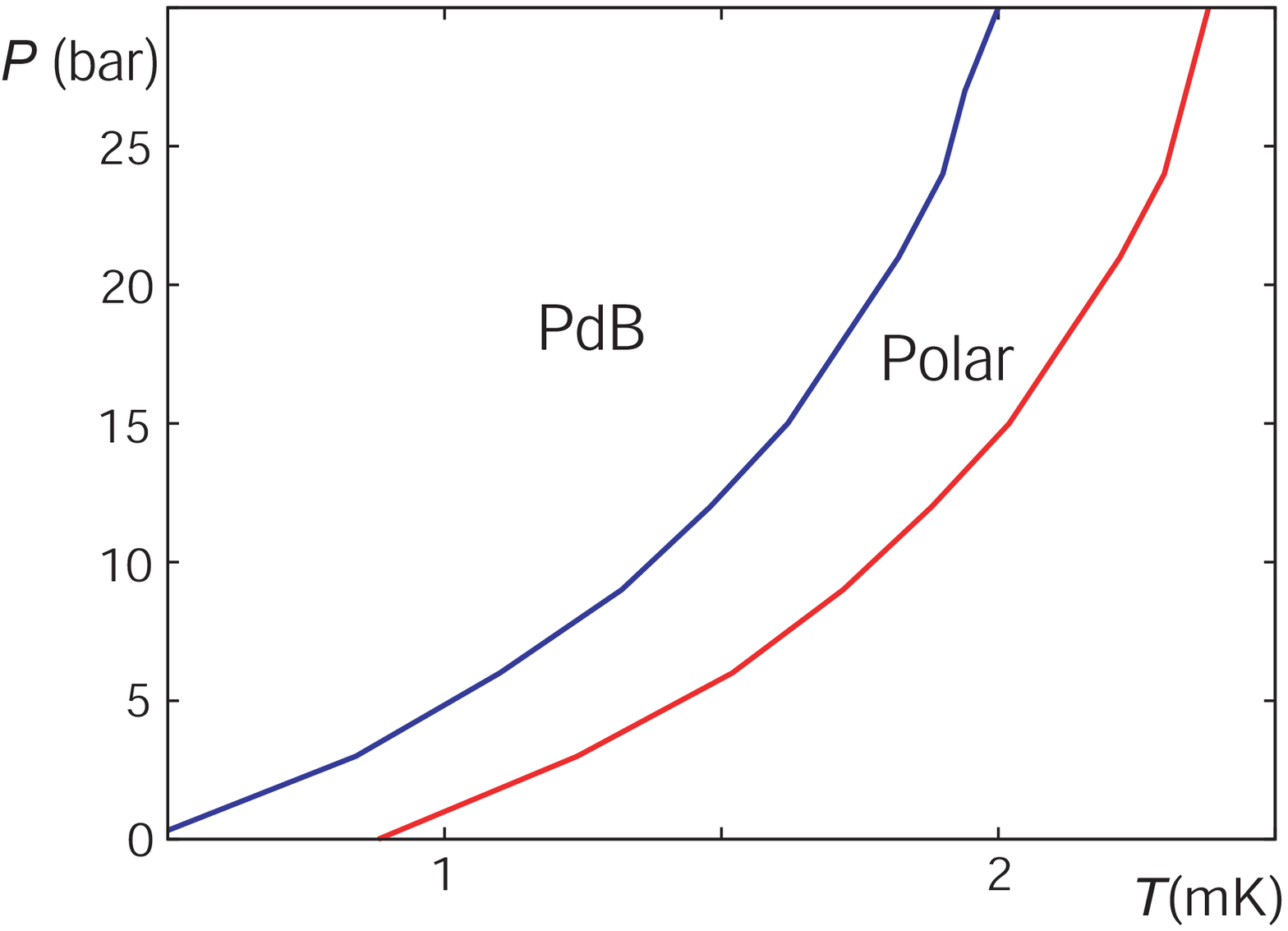}}
{\includegraphics[scale=0.3]{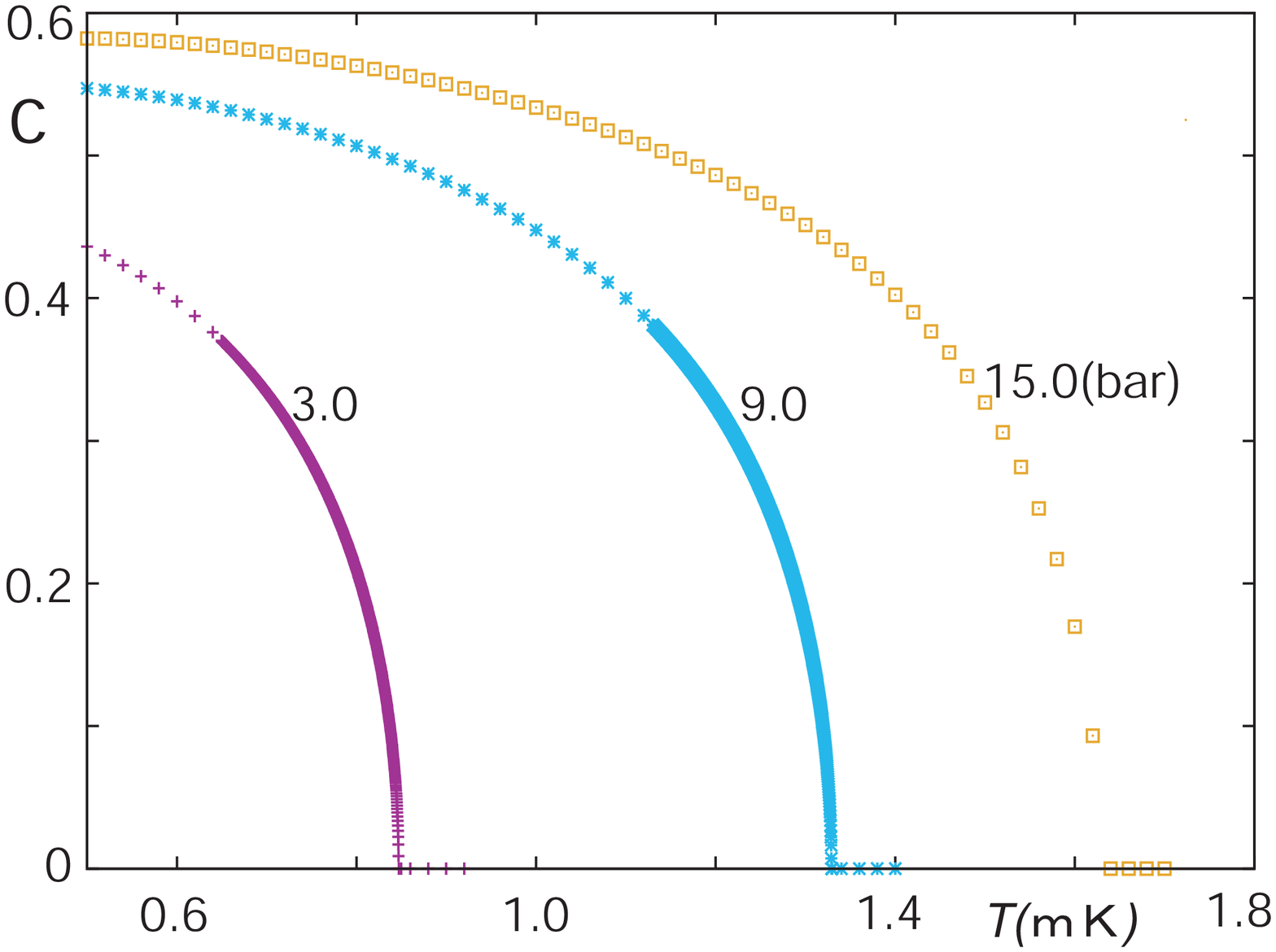}}
\caption{Numerical results for $\delta=4.4$. The pressure $P$ v.s. temperature $T$ phase diagram (a) in which the red curve is the superfluid transition curve $T_c(P)$ between the polar and normal phases, and the blue one is $T_{\rm PB}(P)$. The figure (b) shows the temperature dependence of the "order parameter" $c$ in the PdB phase at the pressures $P=3.0$ (bottom), $9.0$, and $15.0$ (top) (bar).}
\label{s:fig:4.41}
\end{figure}

\begin{figure}[t]
{\includegraphics[scale=0.35]{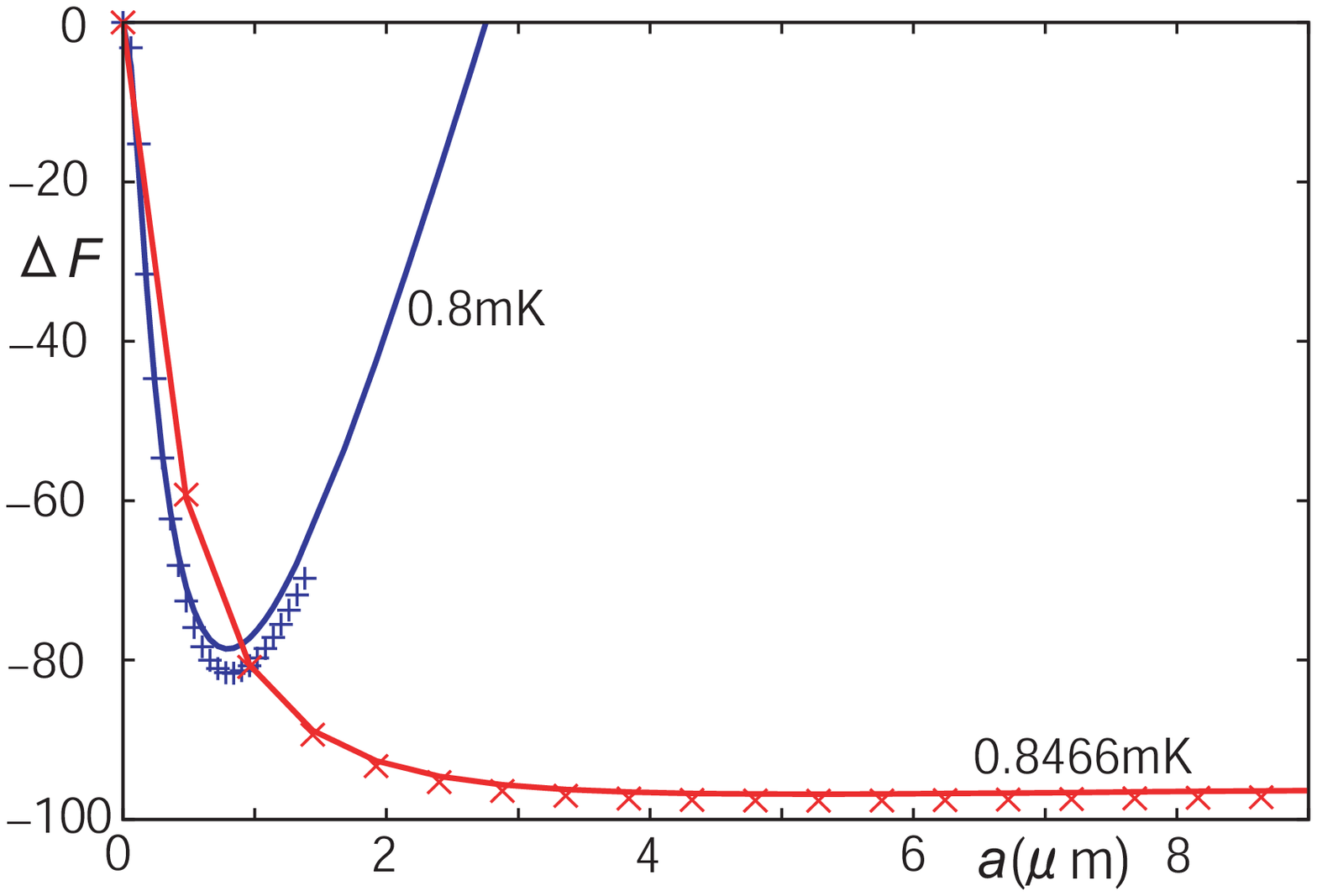}}
{\includegraphics[scale=0.3]{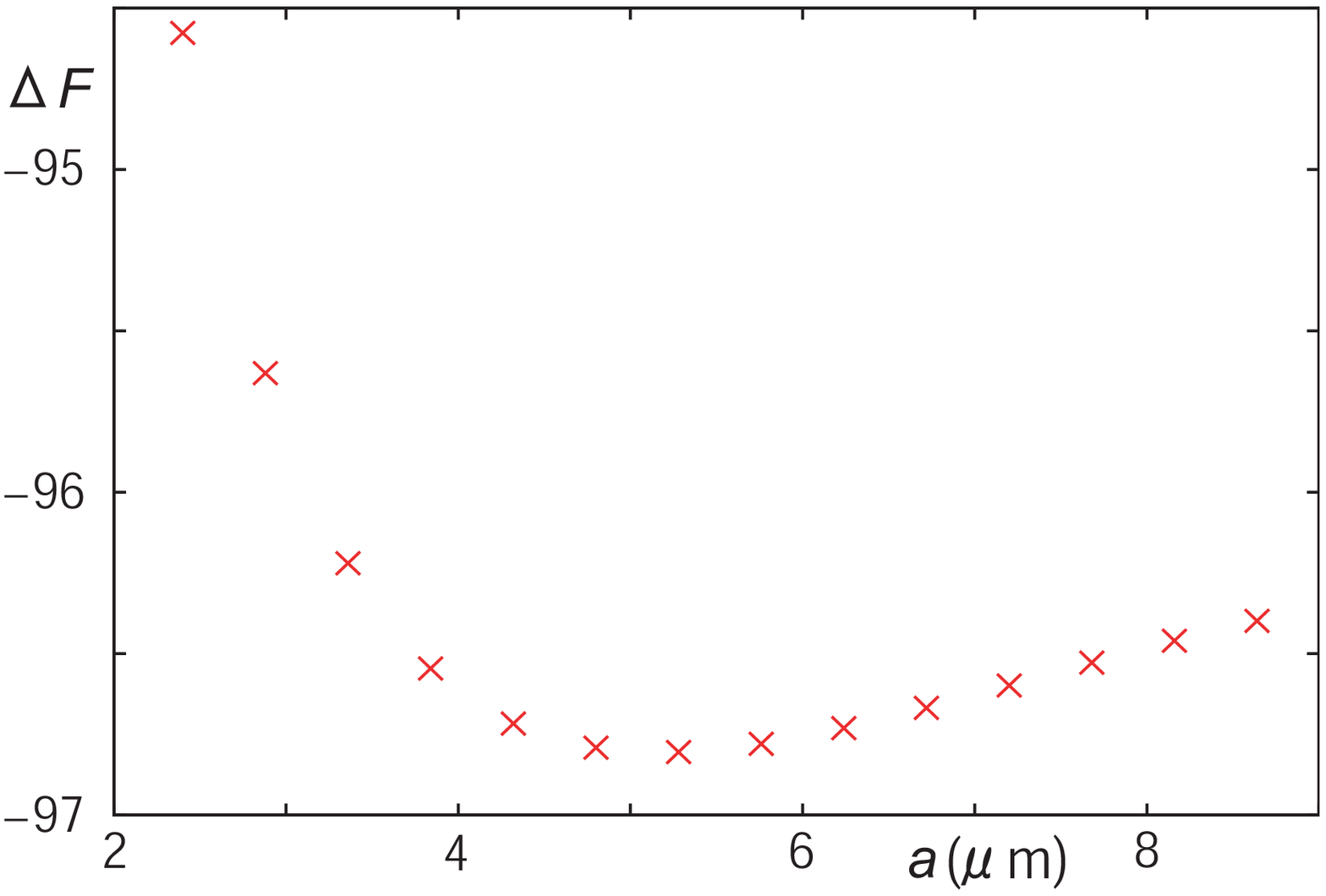}}
\caption{Numerical results for $\delta=4.4$ at $3$(bar). In the figure (a), the $a$ v.s. $\Delta F(a)$ curves are expressed by the blue solid curve at $0.8$(mK) and by the red solid curve at $0.8466$(mK) close to $T_{\rm PB}$, and they have $a_m=0.78$ ($\mu$m) and $a_m=5.28$ ($\mu$m), respectively. For comparison, the results (plus and crossed symbols) in the case with additional vertex corrections (see the text) are also shown. The figure (b) is the extended view of the red symbols around $a=5.28$ in (a). }
\label{s:fig:4.41}
\end{figure}

As Fig.4(a) shows, the polar phase region in this $\delta = 0.05$ case is extremely narrow, and $T_{\rm PB}$ and $T_c$ curves are quite close to each other. In Fig.4(b), the dependence of the vortex core energy $\Delta F$ on the initial value $2a$ of the half-core pair size is presented for the two values of $c(T)$. As Fig.6(b) shows, the parameter $c$ playing the role of the order parameter in the PdB phase grows upon cooling. The $2a$ value minimizing $\Delta F$ corresponds to the half-core pair size $2a_m$ to be realized. Closer to the phase boundary $T_{\rm PB}$ at which $c$ vanishes, the $a_m$ value becomes larger as suggested by eq.(28), Further, as the solid curve in Fig.4(b) shows, the conventional GL free energy with no FL correction term eq.(29) results in a smaller size $2 a_m=0.48$($\mu$m) of the half-core pair \cite{Th2,NI2}. Such a correlation-induced growth of the half-core pair size has been pointed out elsewhere \cite{Th2,NI2}

Figure 5 (a) and (b) show spatial variations of $A_{\mu j}$ on sweeping along the $x$ and $y$ axis, respectively, for $c=0.2$ and $\delta=0.05$. Broadly speaking, the midpoint of $A_{yz}$ and $A_{zz}$ curves correspond to the position of the half-core, i.e., $x=a_m$. In the isotropic case where $c^2=2/3$ irrespective of the temperature, $|A_{yz}|$ and $|A_{zy}|$ at the origin coincide with each other. A large difference between them at $x=0$ appears in Fig.4(b) due to the "anisotropy" value, $c=0.2$. This can be understood from eq.(21) with $\phi_+=\pi$ and $\phi_-=0$. On approaching the vortex center along the $x$-axis, $A_{xx}$ decreases. Nevertheless, $A_{xx}$ seems to be nonvanishing even close to $x=0$. It means that the planar state is realized only in the close vicinity of the origin. Further, the width $\xi_w$ indicated in Fig.2(b), i.e., range of $y$ over which $A_{xx}$ linearly decreases is large ($\simeq 0.3$) as Fig.5(b) shows. These behaviors of $A_{xx}$ imply that, in spite of a substantial size $2 a_m$ of the half-core pair (see Table 1), the planar string (wall) expected in the London description in sec.III is ill-defined when $\delta=0.05$. In fact, the $a$-dependence of $\Delta F(a)$ in $a > a_m$ in Fig.4(b) seems to be different from the expected linear behavior in $a$. Therefore, the half-core structure of the double-core vortex cannot be identified with a HQV pair in the case of low anisotropy such that $\delta=0.05$. 

\begin{figure}[t]
{\includegraphics[scale=0.35]{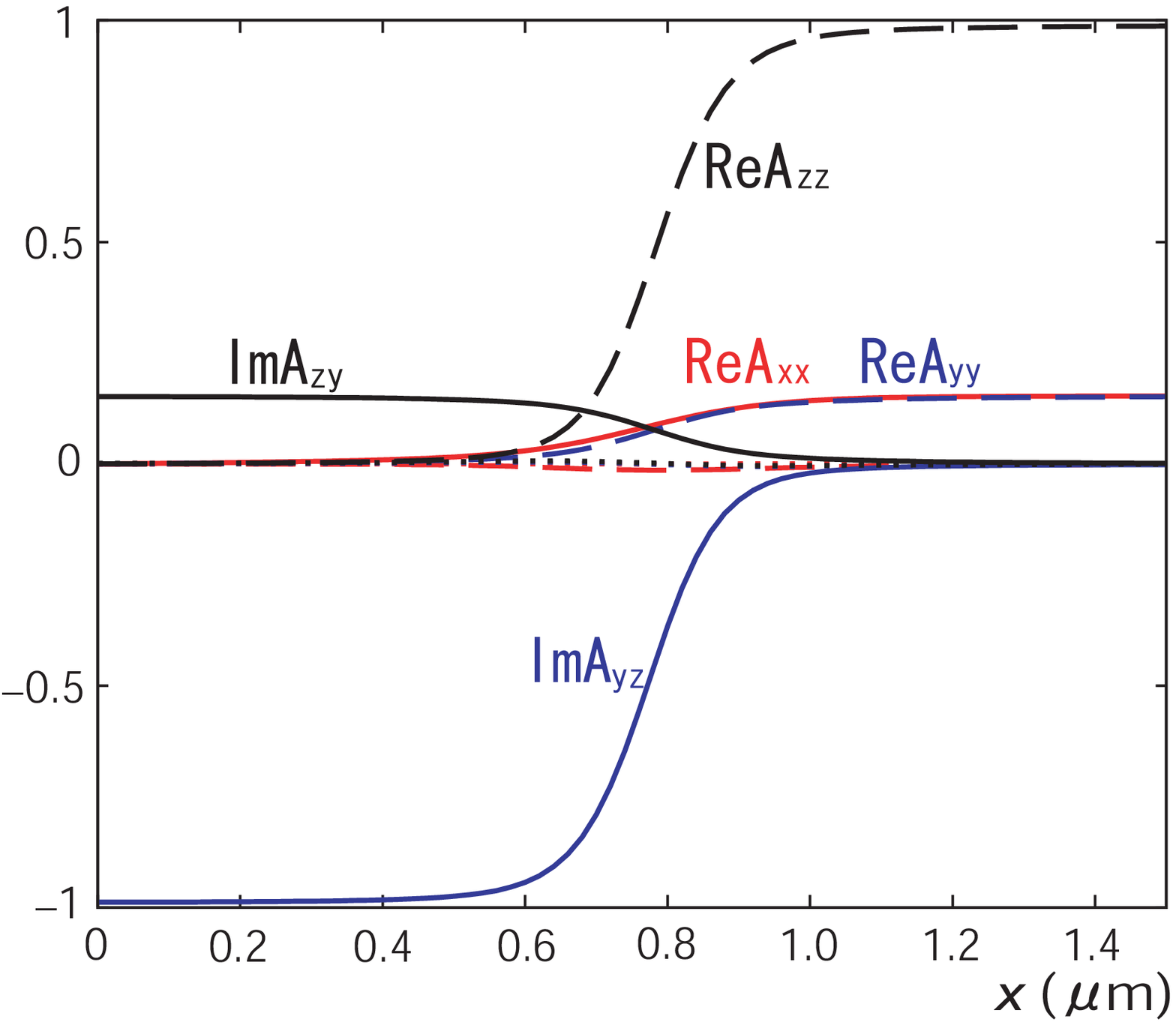}}
{\includegraphics[scale=0.35]{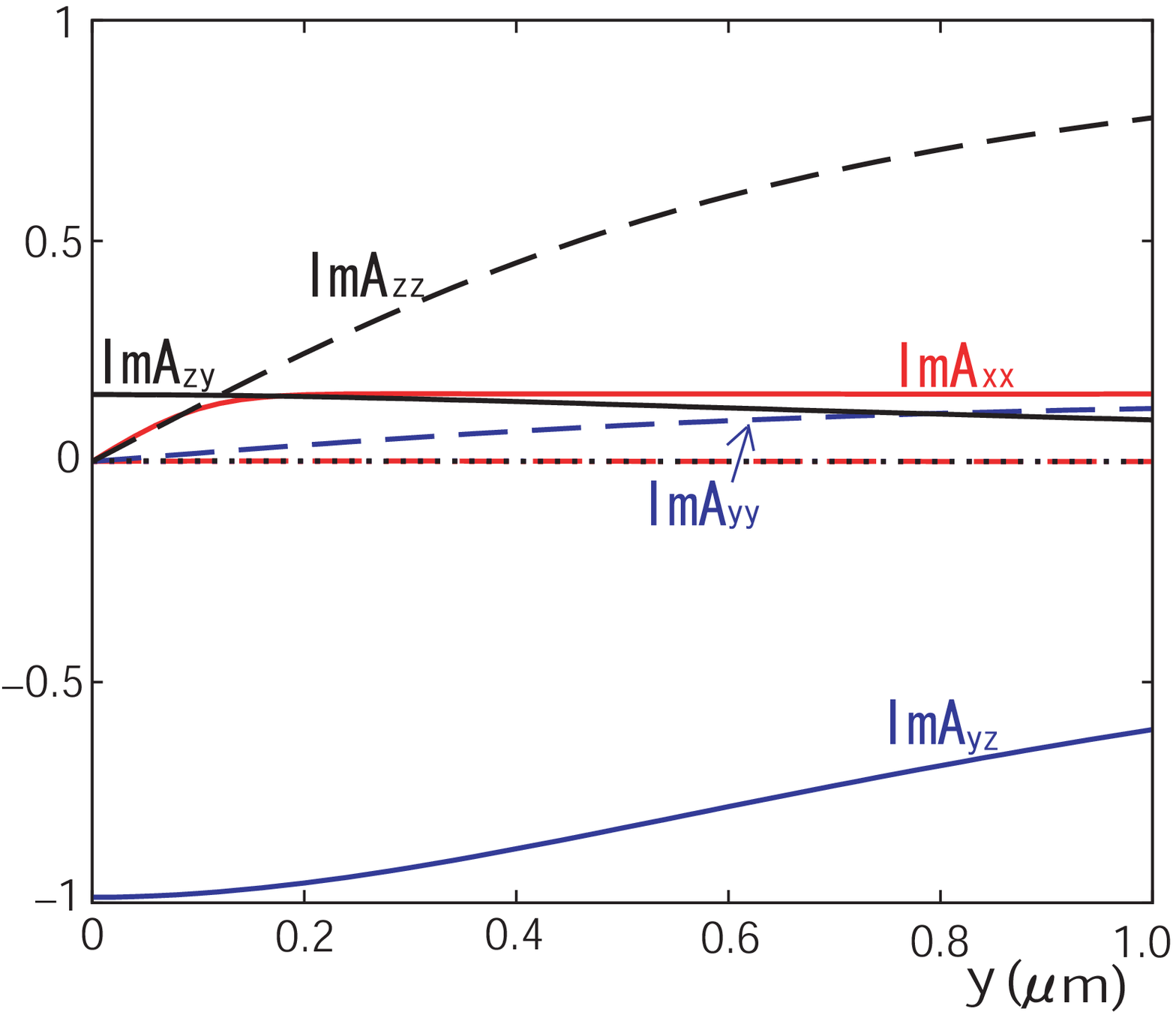}}
\caption{The figure (a) expresses the data, at $a=a_m=0.78$ of the blue solid curve in Fig.7(a), of spatial variations of $A_{\mu j}$ components on sweeping along the $x$-axis and at $y=0$. The figure (b) corresponds to (a) on sweeping $y$ and at $x=0$. Here, the vortex center is at ($0$, $0$). 
}
\label{s:fig:4.42}
\end{figure}

Next, the corresponding results in a case with a moderately strong anisotropy, $\delta=4.4$, are presented in Figs.6 and 7. The $P$-$T$ phase diagram and the temperature dependence of the order parameter $c$ of the PdB phase are given in Fig.6(a) and (b), respectively. Figure 6(a) shows that a moderately wide region of the polar phase is realized in this case. As pointed out elsewhere \cite{Eltsov19,IT19,VVD3}, the superfluid transition (right solid) curve is not changed notably depending on the "impurity" strength $\tau^{-1}$, while the $T_{\rm PB}(P)$ (left solid) curve is sensitive to $\tau^{-1}$, and a slight increase of $\tau^{-1}$ remarkably broadens the polar phase region. In Fig.7(a) and Fig.8, we focus on the $P=3$(bar) case and on the results at the two temperatures, $0.8466$(mK) at which $c=0.0456$ or $c/c_{\rm M} \simeq 0.094$) and $0.8$(mK) at which $c=0.214$ or $c/c_{\rm M} \simeq 0.439$, where $c_{\rm M}(P)$ is the $c$ value in $T \to 0$ limit at each pressure. 

Figure 8(a) and (b) express the spatial variations of the components of $A_{\mu j}$ at $0.8$ (mK) in $P=3$(bar) when $\delta = 4.4$. Some clear differences between Fig.8(a) and Fig.5(a) are seen. First, in the notation of eq.(24), the following relations are satisfied in Fig.8(a); $\theta \simeq 0$ and $|A_{xx}| \simeq c/\sqrt{2}$ in $x > a_m$, while $\theta \simeq \pi/2$ and $|A_{xx}|=0$ in $0 \leq x < a_m$. Next, the linearly vanishing behavior of $A_{xx}$ (red solid curve) on lowering $|y|$ and at $x=0$ is seen only in a narrow region near the origin so that $\xi_w \simeq 0.1$($\mu$m). In addition, the linear behavior $\Delta F(a) \propto a$ is nicely seen in $a > a_m$ in Fig.7 (a) and (b), implying that the planar string is well-defined and has a length comparable with the size $2 a_m$ of the half-core pair. In fact, Fig.7 (a) and (b) shows that such a linear behavior approximatedly obeys the relation $S(T) c^2 a$ where the $T$-dependent coefficient $S(T)$ slowly increases upon cooling, i.e., a relation consistent with the London result eq.(27). 
Further, except in the vicinity of the half-core, other components of $A_{\mu j}$ than the five nonvanishing ones in eq.(21) can be regarded as being zero. 

Based on these features, in contrast to the $\delta=0.05$ case, the double-core vortex in $\delta=4.4$ is consistent with the description in London limit and can be well regarded as a HQV pair. However, the origin of this consistency with eq.(24) cannot be ascribed merely to the growth of the half-core pair. For instance, the anisotropic growth of the double vortex core also may occur due to an enhanced rigidity. In such a situation expected to occur in {\it isotropic} aerogel \cite{NI2}, the growth of the half-core pair is accompanied by the corresponding {\it enhancement} of the components of $A_{\mu j}$ which are zero in eq.(21) \cite{NI2}, contrary to the feature seen in Fig.8(a) and (b). The reason why the double core vortex in the PdB phase in such a moderately strong anisotropic case is well described by the London limit seems to consist in the simple structure of the HQV in $c \to 0$ limit, i.e., in the polar phase. As shown in Ref.\cite{NI1}, the spatial variations of the order parameter are surprisingly simple and are well represented by eq.(24) with $c=0$ and $A_{xx}=0$ except in the close vicinity of each HQV. A smaller $c$-value effectively implying a stronger anisotropy leads to a structure closer to that in the London limit. 

Further, in Fig.7(a), we have also presented the $\Delta F(a)$ v.s. $a$ curves (plus and crossed symbols) in the case where the O($|\Delta|^4$) gradient energy includes all of the vertex corrections accompanied by $C$, $B$, and $D$ in eq.(7). The deviation from the case (crossed symbols) with no such vertex corrections in the O($|\Delta|^4$) gradient energy is negligibly small, and the resulting $a_m$ value is not affected much by including such vertex corrections. Therefore, we judge that the neglect of the vertex corrections to the O($|\Delta|^4$) gradient terms, mentioned in the beginning of this section, is valid in all of other results presented here. 

Figure 9 (a) and (b) express spatial variations of $A_{\mu,j}$ for a much stronger anisotropy, $\delta=300$. Here, to examine the anisotropy dependences of the obtained results at the same pressure, we have compared the results at the same value of the normalized $c$-value, $c(T)/c_{\rm M}$, with each other. Note that $c(T)/c_{\rm M}$ measures the distance from the polar to PdB transition line $T_{\rm PB}(P)$. Surprisingly, Fig.8 and Fig.9 are qualitatively similar to each other, suggesting that the $\delta=4.4$ case already enters the limit of the strong anisotropy. Nevertheless, the resulting size $a_m$ of the HQV pair minimizing the energy at the same pressure and at the same $c/c_{\rm M}$-value increases with increasing the anisotropy $\delta$ (see Table I to be given later). 

\begin{figure}[t]
{\includegraphics[scale=0.35]{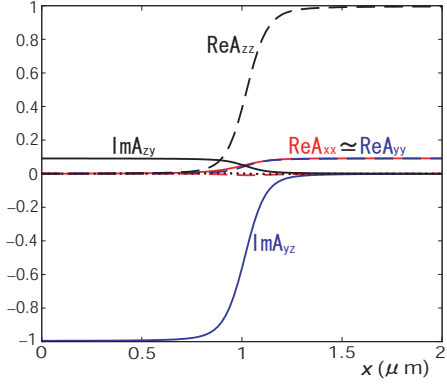}}
{\includegraphics[scale=0.35]{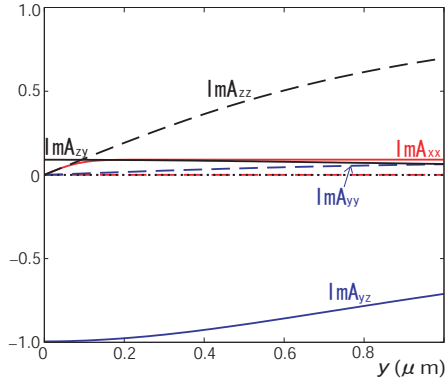}}
\caption{Spatial variations of $A_{\mu j}$ at $0.61$(mK) below $T_{\rm PB}$, where $c=0.127$, at $3$(bar) for $\delta=300$ obtained on sweeping (a) along the $x$ axis and (b) along the $y$ axis, respectively. 
}
\label{s:fig:300}
\end{figure}
In Table I, examples of our numerical data of the resulting $a_m$ value (the half of the HQV pair size) are presented as a function of $\delta$ and $P$. Regarding the pressure dependence, the results in $P=3$ (bar) and $9$ (bar) at a fixed $c/c_{\rm M}$ value indicate that the resulting $a_m$ becomes smaller with increasing $P$. However, it should be noted that, under a fixed $c/c_{\rm M}$ value, an increase of $P$ corresponds to an increase of $T$ according to Fig.6(b), and that, as eq.(28) suggests, $a_m$ increase upon cooling through not only $c/c_{\rm M}$ but, e.g., the gap amplitude $|\Delta|(T)$.

\begin{table}[t]
\begin{center}
\small
\begin{tabular}{cccccccccc} \hline
    $\delta$ & 0.05 & 0.05 & 4.4 & 4.4 & 4.4 & 300& 300 & \\ \hline
    $P$(bar) & 9 & 9 & 9 & 3 & 3 & 3 & 3 & \\
    $T$(mK) & 1.447 & 1.469 & 1.28 & 0.8 & 0.8466 & 0.61 & 0.6396 & \\  
    $c/c_{\rm M}$ & 0.628 & 0.251 & 0.391 & 0.439 & 0.094 & 0.363 & 0.157 & \\
    $a_m$($\mu$m) & 0.36 & 0.6 & 0.54 & 0.78 & 5.28 & 1.02 & 6.24 & \\ 
\hline
  \end{tabular}
\label{amdata}
\caption{Resulting $a_m$ values at different temperatures for various $\delta$-values at $3$ and $9$(bar). The $c/c_{\rm M}$-value in each case is also shown, where $c_{\rm M}=c(T=0)$. }
\end{center}
\end{table}

On the other hand, the $c$ dependence itself of $a_m$ is not necessarily understood when comparing with the expectation from the London limit. In fact, the $c$ dependence of $a_m$ is much weaker than that suggested by eq.(28), and the $a_m$ value increases only weakly with vanishing $c$. Judging from the fact that the linear $a$ behavior of $\Delta F$ is seen in Fig.7 (b), this discrepancy deos not seem to be due to the smallness of the system size. On the other hand, the logarithmic behavior of eq.(26), which has been nicely verified in the polar phase \cite{NI1}, is masked in the present case by the linear behavior eq.(28). Hence, we cannot clarify whether the contribution corresponding to eq.(26) is satisfied or not in the present case. 
\vspace{5mm}

\section{Summary}

In this work, we have numerically examined the stability of a HQV pair in the PdB phase of the superfluid $^3$He in a strongly anisotropic aerogel by assuming the weak coupling approximation and based on the hypothesis that the double-core vortex in the bulk B phase corresponds to the HQV pair in the PdB phase. Due to the weak coupling approximation, the presence of the PdA phase in real systems is neglected, and the transition between the PdB and the polar phases becomes inevitably continuous in the present analysis. However, such a continuous transition is found at low enough pressures in real systems \cite{Eltsov2}, and in this 
sense the present results may be directly applicable to the experimental 
situations. 

Our main result in the present work is that the double-core vortex \cite{Th87,SVreview} in the PdB phase under a strong anisotropy can be regarded as a HQV pair described in the London limit. This is a reflection of the fact that the HQV in the pure polar phase is well described in the London limit \cite{NI1}. 
\begin{widetext}
\begin{figure}[t]
{\includegraphics[scale=0.35]{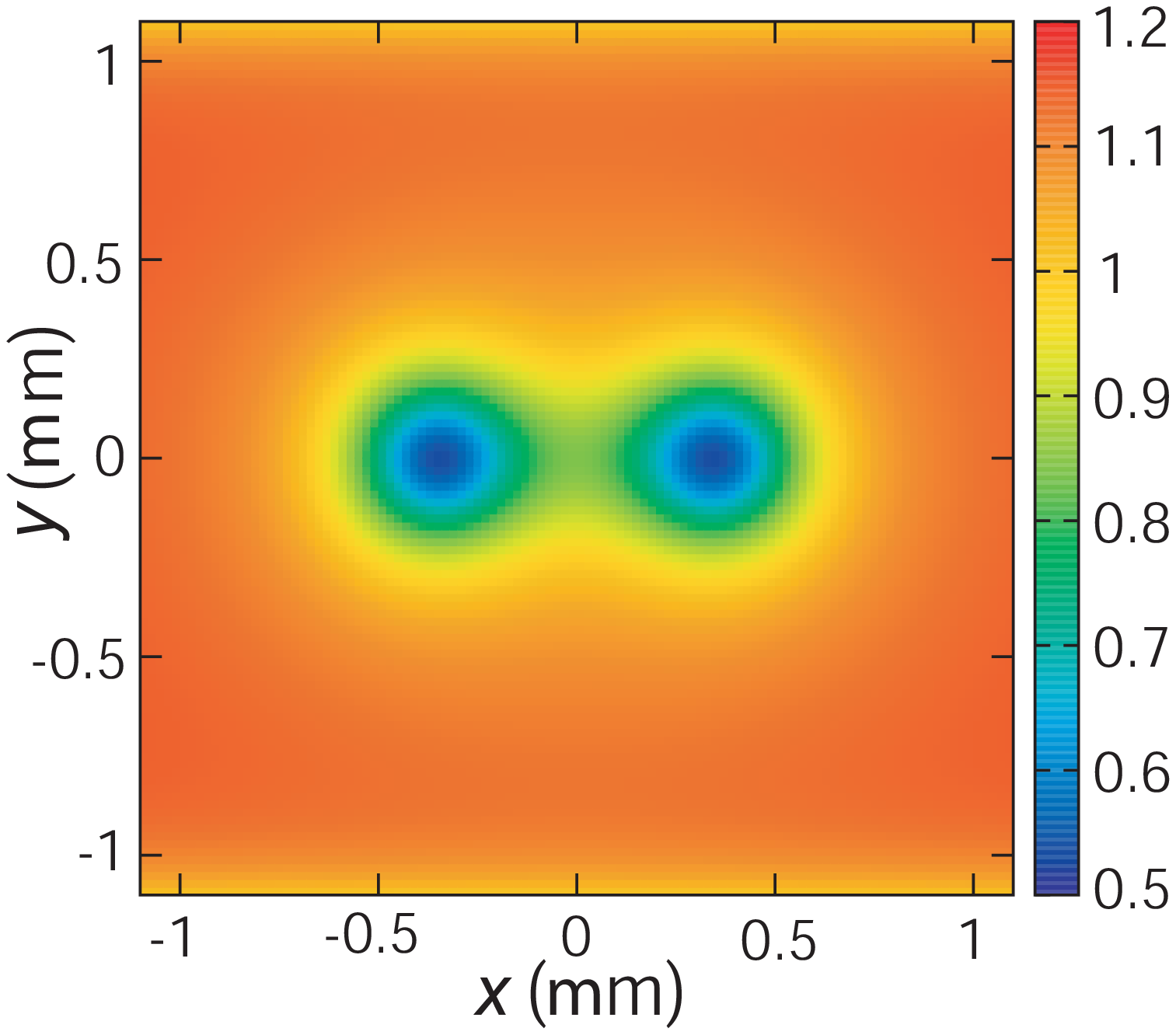}}
{\includegraphics[scale=0.35]{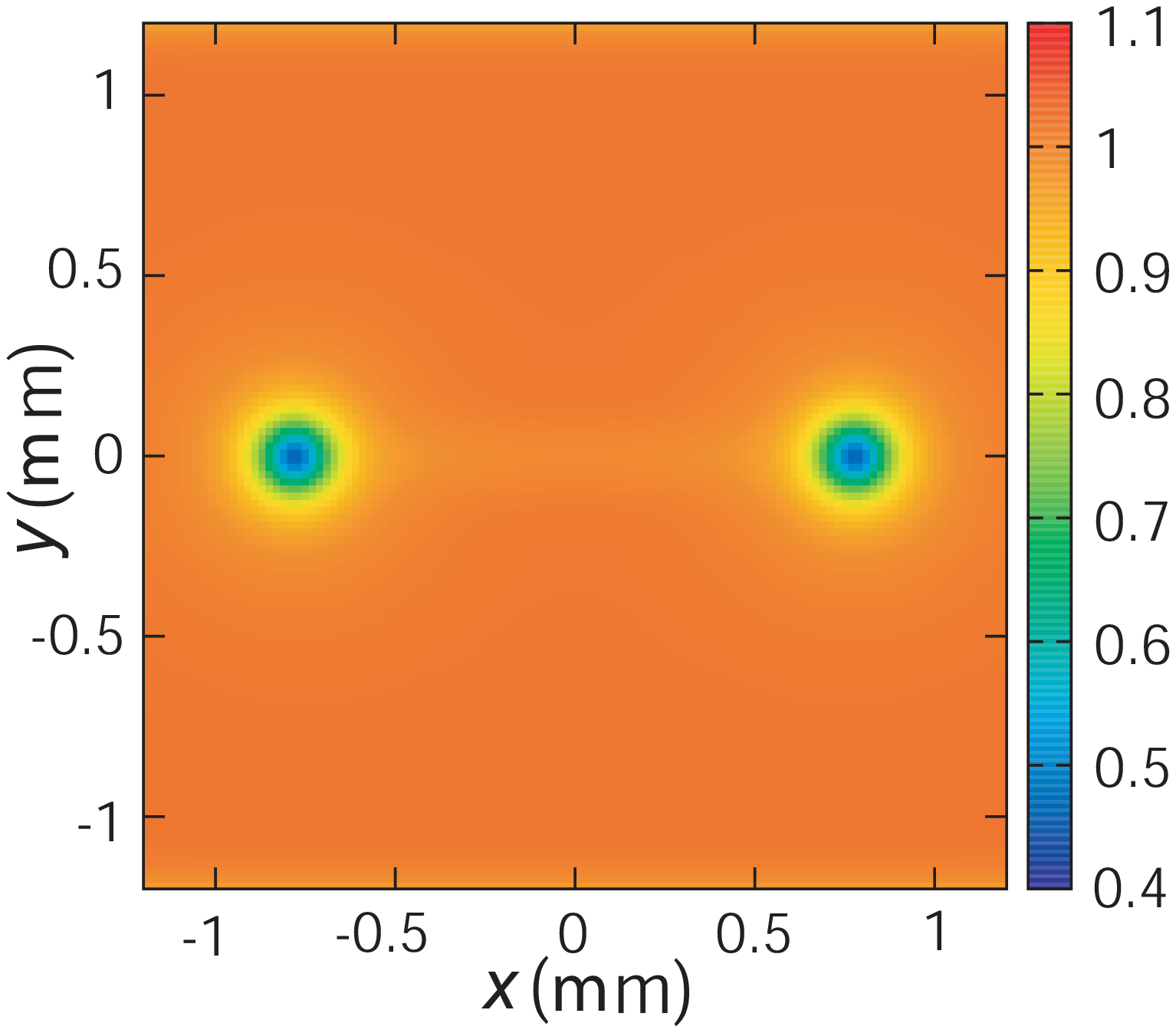}}
{\includegraphics[scale=0.35]{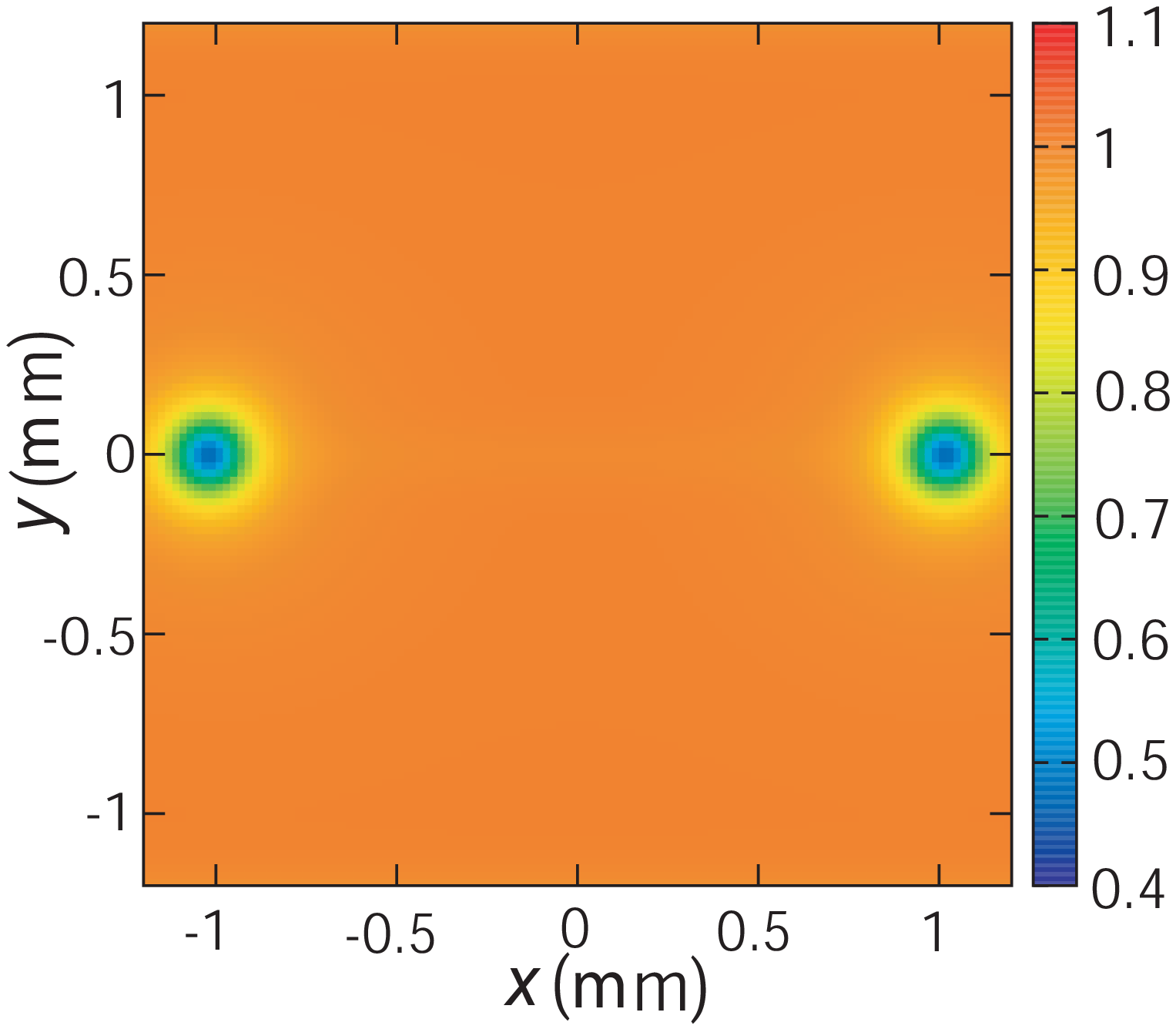}}
\caption{Spatial profiles of the squared amplitude of the order parameter, i.e., $\sum_{\mu, j} |A_{\mu j}|^2$ in the $x$-$y$ plane in the cases of $\delta=0.05$ (top), $\delta=4.4$ (middle), 
and $\delta=300$ (bottom). The $c$-value is $0.5$ in the top figure. The middle figure corresponds to Fig.8, while the bottom one corresponds to Fig.9. A dim string representing the planar wall connecting the two HQVs is seen in the middle and bottom figures, and the core size of each HQV is found to become smaller with increasing $\delta$. }
\label{s:fig:300}
\end{figure}
\end{widetext}

In Fig.10, the color maps expressing the spatial distribution of the squared amplitude of the order parameter $\sum_{\mu,j} |A_{\mu j}|^2$ are presented for different anisotropy values. In $\delta = 0.05$ (top figure), the planar string (wall) is ill-defined, and the two half-cores, i.e., the two HQVs, are connected rather by a 
broad region with a diminished amplitude of the order parameter. In contrast, with increasing the anisotropy, the two HQVs tend to be connected only by a thin planar string on which the amplitude of the order parameter is faintly diminished, and each core of HQVs becomes sharper. In this manner, the description in the London limit becomes better for larger anisotropies. 

\begin{widetext}
\begin{table}[htb]
\begin{center}
\small
\begin{tabular}{cccccccccc} \hline
    $\delta$ & 0.05 & 0.05 & 4.4 & 4.4 & 4.4 & 4.4 & 300& 300 & \\ \hline
    $P$(bar) & 9 & 9 & 9 & 9 & 3 & 3 & 3 & 3 & \\
    $T$(mK) & 1.447 & 1.469 & 1.28 & 1.33 & 0.8 & 0.8466 & 0.61 & 0.6396 & \\  
    $c/c_{\rm M}$ & 0.628 & 0.251 & 0.391 & 0.058 & 0.439 & 0.094 & 0.363 & 0.157 & \\
    $\xi_w$($\mu$m) & 0.2 & 0.286 & 0.073 & 0.121 & 0.105 & 0.226 & 0.092 & 0.299 
& \\
\hline
  \end{tabular}
\label{amdata}
\caption{Resulting thicness $\xi_w$ of the planar (Kibble) wall obtained based on the definition given in the text at different temperatures for various $\delta$-values at $3$ and $9$(bar). }
\end{center}
\end{table}
\end{widetext}
Here, the present result will be compared with the experimental result in Ref.\cite{Eltsov}, where the HQV pairs in the PdB phase have been detected. To understand the structure of the resulting HQV pair better, we have estimated the width of the planar string $\xi_w$ (see Fig.2(b)). Broadly speaking, this length is etimated by comparing the mass term and the gradient one of O($A_{xx}^2$) with each other to be of the order of $\xi(T)/c$ \cite{Volovik90,Eltsov}. From the numerical data, $\xi_w$ will be defined by assuming that the $y$ dependence of $A_{xx}$ close to the origin is approximated by $c \, {\rm tanh}(y/\xi_w)/\sqrt{2}$ \cite{Eltsov} (see also eq.(21)), As shown in Table II, $\xi_w$ indeed grows with decreasing $c/c_{\rm max}$, though the $c$ dependence is apparently weaker compared with the relation mentioned above. The deviation from the $c^{-1}$ dependence seems to be resolved by noting that both $c$ and $\xi(T)$ are $T$-dependent. On the other hand, the $\delta$ dependence of $\xi_w$ is not anticipated easily and is found only through the present numerical analysis. Table II suggests that, with increasing $\delta$, the aspect ratio $2 a_m/\xi_w$ becomes large enough to make the planar string a rigid and well defined object. We note that this ratio is inversely proportional to $c$, reflecting the proximity to the polar phase in which the HQV pair is infinitely long with no dipole energy neglected in the present analysis. 
 
The results on the vortex energy shown in Figs.4 and 7 imply together with the data in Table I that the size of the HQV pair, $a_m$, is highly sensitive to the $c$-value. On the other hand, according to Ref.\cite{Eltsov}, huge HQV pairs have appeared in the PdB phase in spite of a reasonable $T$ dependence of $c$ corresponding to $\sqrt{2} q$ there (see Supplementary Fig.5 in Ref.\cite{Eltsov} which is qualitatively comparable with Fig.6(b)). 
It is an evidence of the presence of a strong pinning effect in real systems supporting the huge HQV pair in the nematic aerogels \cite{Eltsov}. 

In the experiment under rotation \cite{Autti}, the rotation axis has been fixed to the anisotropy (polar) axis, which is the $z$-direction in our notation. Further, as mentioned above, the vortices created under a rapid quench \cite{Eltsov} are also pinned along the polar axis because the mean free path for the quasiparticles is the 
longest in this direction. Thus, it is possible that, if rotating the aerogel with a rotation axis perpendicular to the polar axis, a pinning of the resulting vortices to the aerogel structure amy be avoided. Then, the shrinkage of the HQV pair upon cooling might be observed in such a situation. There is another motivation regarding a study of HQVs extending along a direction perpendicular to the polar axis. Recently, NMR measurements for $^3$He in a nematic aerogel squeezed by $30$ percent in a direction perpendicular to the polar axis have been reported \cite{VVD3}. There, it has been found that the $l$-vector in the chiral PdA phase is largely directed along the squeezed direction. In this situation, the Majorana fermions may remain stable \cite{Majorana} in the core of a HQV in the PdA phase. For these reasons, it will be valuable to extend the present study on HQVs to the situation with the vortex axis perpendicular to the polar axis.

\section{Appendix A}

In this Appendix, details of the pairing vertex correction due to the impurity scattering and of the coefficient of each term in the resulting GL free energy are explained. 

The impurity scattering potential does not carry the Matsubara frequency, and consequently, the corresponding self energy term can be incorporated through the replacement of the Matsubara frequency $|\varepsilon|$ with 
\begin{eqnarray}
|{\tilde \varepsilon}_p| &=& |\varepsilon| + \frac{1}{2 \tau} \langle w(p - p') \rangle_{p'} \nonumber \\ 
&=& |\varepsilon| + \frac{1 + (\delta^{-1/2} - 1) \theta(1-\delta)}{4 \tau} ( \, {\rm tan}^{-1}(\delta^{1/2}(1 -  p)) \nonumber \\
&+& {\rm tan}^{-1}(\delta^{1/2}(1 + p)) \, )
\end{eqnarray}
($|p| < 1$), where $p={\bf p}\cdot{\hat z}/p_{\rm F}$, $\langle \,\,\, \rangle_{p}$ implies the average over the polar angle ${\rm cos}^{-1}(p)$. 

The coefficients composing the vertex part $\Lambda$ are given in the form 
\begin{widetext}
\begin{equation}
\begin{pmatrix}
B_0 & D_0 \cr \Delta B & \Delta D \cr
\end{pmatrix} 
= \frac{1}{2} 
\begin{pmatrix} 
1 - I_{d11}& -\delta^{-1}(I_{d10} - I_{d11}) \cr - \delta(3I_{d12} - 4I_{d13}) & 1 - 3I_{d11}+7I_{d12}-4I_{d13} \cr  
\end{pmatrix}
^{-1} 
\begin{pmatrix}
e_1 & \,\,\, -I_{d21}+ 2 \delta^{-1} C_0(I_{d20}-I_{d21}) -e_1 \cr e_2 
 & \,\,\, 2C_0(3I_{d21}-7I_{d22}+4I_{d23})- e_2 \cr
\end{pmatrix}
\end{equation}
\end{widetext}
where 
\begin{equation}
e_1 = I_{d21} + \delta^{-1}(I_{d21} - I_{d20}), 
\end{equation}
\begin{eqnarray}
e_2 &=& \delta(3I_{d22}-4I_{d23}) \nonumber \\
&-& 3I_{d21}+7I_{d22}-4I_{d23}, 
\end{eqnarray}
and  
\begin{eqnarray}
C_0 &=& \frac{1}{d}, \nonumber \\
C_{21} &=& \frac{-I_{d31}+I_{d32} + \delta^{-1}(I_{d30} - 2I_{d31}+I_{d32})}{d^2}, \nonumber \\
C_{1z} &=& 2 d C_{21} - \frac{2}{d}(B_0(I_{d21}-I_{d22}) \nonumber \\ 
&+& \Delta B (I_{d20} - 2I_{d21}+I_{d22})), 
\end{eqnarray}

\begin{eqnarray}
C_{2z} &=& \frac{1}{d}((2+C_0)(I_{d31}-I_{d32}) - 2D_0(I_{d21}-I_{d22})) \nonumber \\
&-& \frac{1}{d \delta} ((2+3C_0)(I_{d30}-2I_{d31}+I_{d32}) \nonumber \\
&+& 2 \Delta D(I_{d20}-2I_{d21}+I_{d22})). 
\end{eqnarray}
Further, 
\begin{eqnarray}
d &=& 1 - 2(I_{d11}-I_{d12}), \nonumber \\
I_{dmn} &=& \biggl\langle \frac{1+(\sqrt{\delta} - 1)\theta(\delta -1)}{(2|{\tilde \varepsilon}_p)^m \tau (1 + \delta p^2)^n} \biggr\rangle_p. 
\end{eqnarray}

Among these coefficients, the $|\epsilon|$ dependences of $C_0-1$ and $B_0$ are presented in Fig.11(a) and (b), respectively. 
\begin{figure}[t]
{
\includegraphics[scale = 0.35]{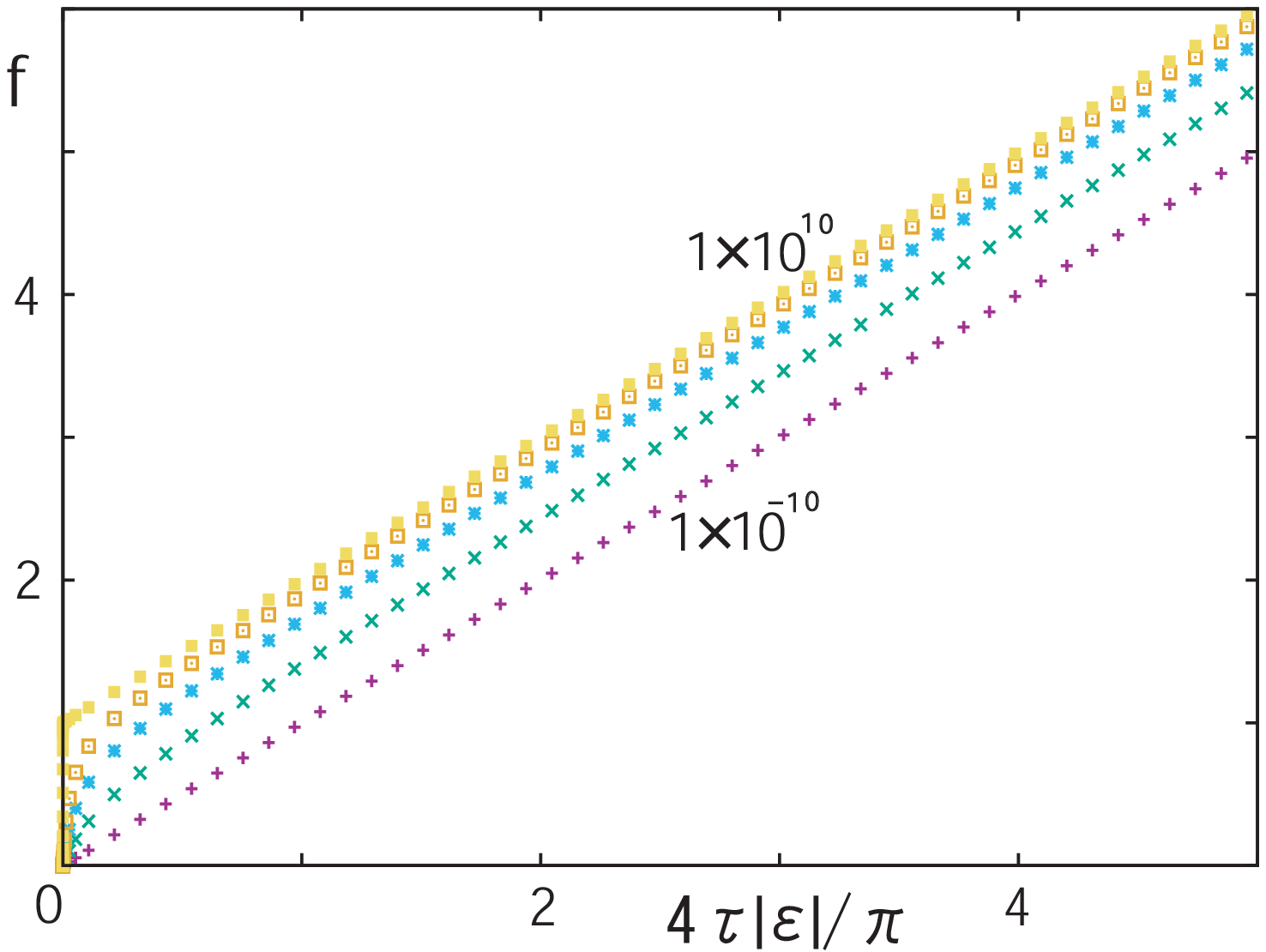}
}
{
\includegraphics[scale = 0.35]{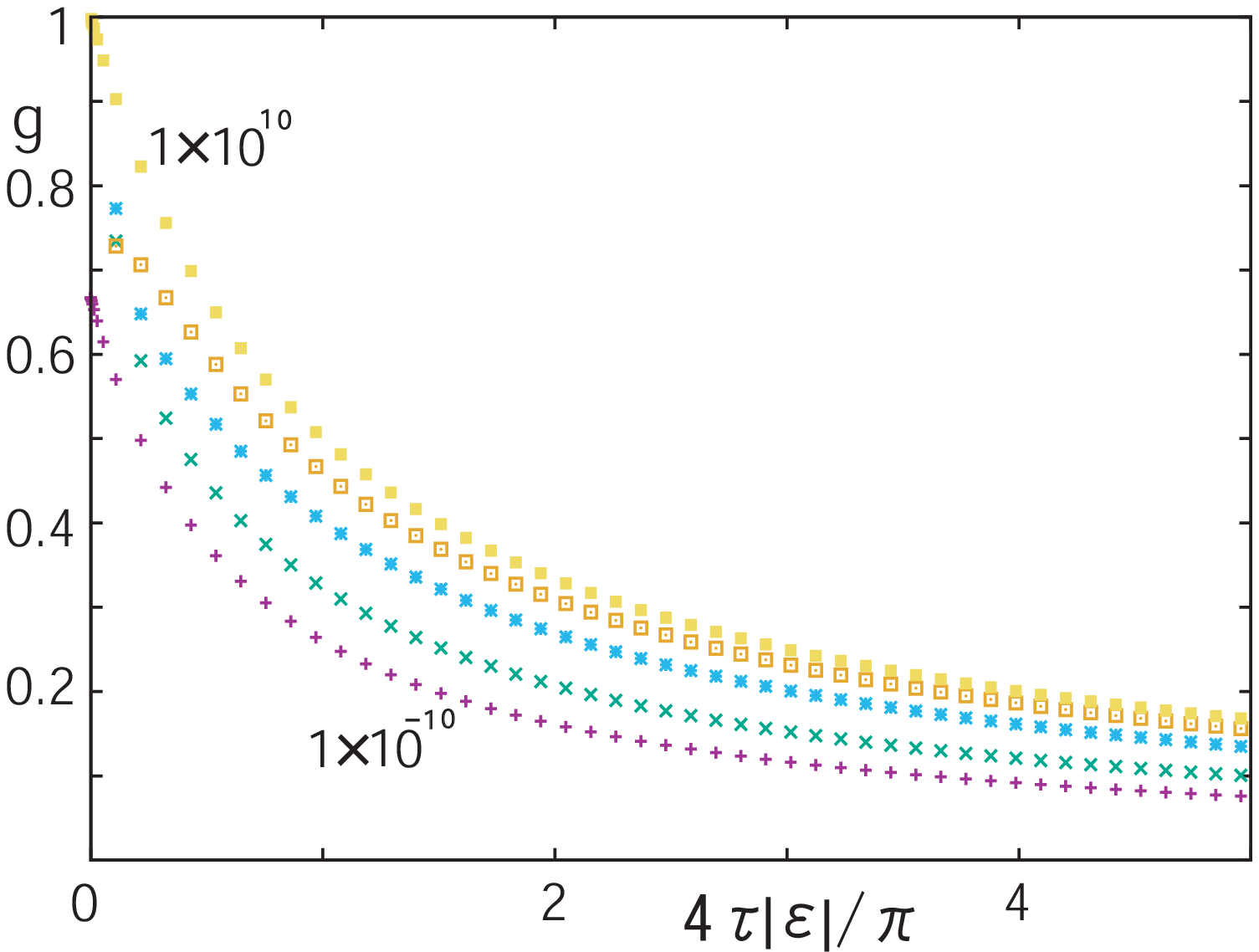}
}
\caption{(a) $f \equiv X C_0$ v.s. $X$ curves for $\delta=1 \times 10^{10}$ (top), $300$, $30$, $4.4$, and $1 \times 10^{-10}$ (bottom), where $X=4 \tau |\varepsilon|/\pi$. Note that $f$ obeys $1+X$ in $\delta \to \infty$, while it approaches $X$ in $\delta \to 0$ limit. (b) $g \equiv \pi X B_0/\tau$ v.s. $X$ curves corresponding to those in (a). In $\delta \to \infty$, $g$ obeys $1/(1+X)$, while it approaches $2/[3(1+\pi X/2)]$ in $\delta \to 0$ in agreement with eq.(9). 
}
\label{s:fig:HQV}
\end{figure}
Here, $f=X C_0$, $g=\pi X B_0/\tau$, and $X=4 \tau |\varepsilon|/\pi$. Broadly speaking, the functions $f$ and $g$ decrease for smaller $\delta$-values. Since, more or less, we focus on the temperature range in which $2 \tau |\varepsilon| \gg 1$, any impurity-induced vertex corrections become negligible in our numerical analysis. 

Next, the coefficients in the GL free energy are given by 

\begin{eqnarray}
\alpha &=& \frac{1}{3} N(0) \biggl[ {\rm ln}\biggl(\frac{T}{T_{c0}} \biggr) + 2 \pi T \sum_{\varepsilon > 0} \biggl( \frac{1}{|\varepsilon|} - \frac{3}{2} ( I_{10} - I_{11} ) \biggr) \biggr], \nonumber \\
\alpha_z &=& \frac{1}{3} N(0) \biggl[ {\rm ln}\biggl(\frac{T}{T_{c0}} \biggr) + 2 \pi T \sum_{\varepsilon > 0} \biggl( \frac{1}{|\varepsilon|} - 3  I_{11} C_0 \biggr) \biggr], 
\end{eqnarray}
where 
\begin{equation}
I_{mn} = \biggl\langle \frac{p^{2n}}{|{\tilde \varepsilon}_p|^m} \biggr\rangle_p, 
\end{equation}
\begin{widetext}
\begin{eqnarray}
\beta_3^{(0)} &=& - 2 \beta_1^{(0)} = \frac{\pi T}{8} N(0) \sum_{\varepsilon > 0} (I_{30} - 2 I_{31} + I_{32}), \nonumber \\
\beta_2^{(0)} &=& \beta_4^{(0)} = - \beta_5^{(0)} = \beta_3^{(0)} - \frac{\pi T}{64 \tau} N(0) \nonumber \\
&\times& \int_{-1}^1 dp_1 \int_{-1}^1 dp_2 \sum_{\varepsilon > 0} \frac{(1 - p_1^2)(1-p_2^2)}{{\tilde \varepsilon}_{p_1}^2 {\tilde \varepsilon}_{p_2}^2 (1+\delta(p_1-p_2)^2)} (1+(\sqrt{\delta} - 1)\theta(\delta-1)), \nonumber \\
\beta_z &=& -\frac{3}{2}(\beta_3^{(0)}+2\beta_3^{(1)}) + \frac{\pi T}{2} N(0) \sum_{\varepsilon > 0} C_0^4 I_{32} - \frac{\pi T}{64 \tau} N(0) \nonumber \\ &\times& \int_{-1}^1 dp_1 \int_{-1}^1 dp_2 \sum_{\varepsilon > 0} \frac{(1 - p_1^2 -2 p_1^2 C_0^2)(1-p_2^2 - 2 p_2^2 C_0^2 )}{{\tilde \varepsilon}_{p_1}^2 {\tilde \varepsilon}_{p_2}^2 (1+\delta(p_1-p_2)^2)}(1+(\sqrt{\delta} - 1)\theta(\delta-1)), \nonumber
\end{eqnarray}

\begin{eqnarray}
\beta_3^{(1)} &=& - 2 \beta_1^{(1)} = -\beta_3^{(0)} + \frac{\pi T}{2} N(0) \sum_{\varepsilon > 0} C_0^2 (I_{31} - I_{32}), \nonumber \\ 
\beta_2^{(1)} &=& \beta_4^{(1)} = - \beta_5^{(1)} = \beta_3^{(1)} + \frac{\pi T}{64 \tau} N(0) \nonumber \\
&\times& \int_{-1}^1 dp_1 \int_{-1}^1 dp_2 \sum_{\varepsilon > 0} \frac{(1 - p_1^2)(1-p_2^2 - 2 p_2^2 C_0^2 )}{{\tilde \varepsilon}_{p_1}^2 {\tilde \varepsilon}_{p_2}^2 (1+\delta(p_1-p_2)^2)}(1+(\sqrt{\delta} - 1)\theta(\delta-1)), 
\end{eqnarray}
\end{widetext}

\begin{equation}
K_2 = \frac{\pi T v^2}{16} N(0) \sum_{\varepsilon > 0} (I_{32} - 2 I_{31}+I_{30}), 
\end{equation}
\begin{equation}
K_3 = \frac{\pi T v^2}{16} N(0) \sum_{\varepsilon > 0} (-5I_{32} + 6 I_{31}-I_{30}), 
\end{equation}
\begin{equation}
K_1 = K_2 + \frac{\pi T v^2}{4} N(0) \sum_{\varepsilon > 0} [(I_{20} - I_{21})B_0 +(I_{21}-I_{22})\Delta B], 
\end{equation}
\begin{equation}
K_4 = -K_2 + \frac{\pi T v^2}{4} N(0) \sum_{\varepsilon > 0} [(I_{31} - I_{32}) C_0 - 8I_{11}C_{21}], 
\end{equation}

\begin{eqnarray}
K_5 &=& 2K_3 + \frac{\pi T v^2}{8} N(0) \sum_{\varepsilon > 0} [(3I_{21} - I_{20})B_0 + (3I_{22}-I_{21})\Delta B 
\nonumber \\
&+& (I_{20} - I_{21})D_0 +(I_{21}-I_{22})\Delta D + 2(I_{31}-I_{32})(C_0-1) \nonumber \\
&-& 8I_{11}C_{1z}], \nonumber \\
K_6 &=& - 7K_3 + \frac{\pi T v^2}{4} N(0) \sum_{\varepsilon > 0} [(3I_{21} - I_{20})D_0 \nonumber \\
&+& (3I_{22}-I_{21})\Delta D + (5I_{32}-3I_{31})(C_0-1)- 8I_{11}C_{2z} \nonumber \\
&+& 3I_{31}-I_{30}].
\end{eqnarray}

\section{Appendix B}

Here, possible effects of the pairing vertex correction, peculiar to the anisotropic scattering, on the O($|\Delta|^4$) gradient energy arising from the Fermi liquid repulsive interaction will be discussed. In our previous work \cite{NI1}, such a vertex correction was not taken into account there by assuming a weak anisotropy.

Since, in the present work, only a straight vortex line extending along ${\hat z}$ is considered, the gradient does not have to include its $z$-component $\partial_z$ in the gradient terms. Then, the only vertex correction in the FL-corrected gradient terms is the factor $C_0-1$ accompanying $A_{\rho z}$ in eq.(29). For instance, the terms including $A_{\mu z}$ in the first line of eq.(29) have to be replaced by 
\begin{widetext}
\begin{eqnarray}
&=& \frac{N(0)}{225} \Gamma_1^s (\pi v)^2 \biggl[ \biggl( T \sum_{\varepsilon > 0} \frac{1}{\varepsilon^3} C_0 \biggr)^2 \biggl[(\nabla\cdot A_\mu)(\nabla \cdot A_\lambda^*)A_{\mu z}^*A_{\lambda z} + (A_\lambda \cdot \nabla)A_{\lambda z}^*(A_\mu^* \cdot \nabla)A_{\mu z} \biggr] \nonumber \\
&+& 
\biggl( T \sum_{\varepsilon > 0} \frac{1}{\varepsilon^3} \biggr) \biggl( T \sum_{\varepsilon > 0} \frac{1}{\varepsilon^3} C_0^2 \biggr) \biggl[ \, \sum_{j=x,y} (\nabla A_{\mu z}) \cdot (\nabla A_{\lambda j}^*)A_{\mu z}^*A_{\lambda j}  + {\rm c.c.} \, \biggr] \nonumber \\
&+& \biggl( T \sum_{\varepsilon > 0} \frac{1}{\varepsilon^3} C_0^2 \biggr)^2 (\nabla A_{\mu z}) \cdot (\nabla A_{\lambda z}^*)A_{\mu z}^*A_{\lambda z} 
\biggr].
\end{eqnarray}
\end{widetext}

As can be seen in Fig.11, however, $C_0(\varepsilon)-1$ remains almost zero irrespective of the $\delta$-value except at low enough values of $2 \pi T \tau$ and is quantitatively negligible in the temperature region where $2 \pi T \tau \gg 1$ is satisfied. Therefore, we can proceed our analysis without incorporating the impurity-induced vertex correction to the pairing process in the FL gradient term even in the limit of strong anisotropy. 

In another gradient terms stemming from the "Gor'kov box", i.e, the ordinary weak-coupling O($|\Delta|^4$) term unaccompanied by a repulsive interaction between quasiparticles, the vertex corrections other than $C_0-1$ are also present. As shown in sec.V of Ref.\cite{NI1}, this weak-coupling diagram does not contribute to the stability of HQVs in the polar and A phases irrespective of how the gradients operate onto the order parameter fields. Further, as explained in relation to Fig.7, the weak coupling O($|\Delta|^4$) term plays only negligible roles for the stability of a HQV-pair occurring in the B phase. 

One of the authors (R.I.) is grateful to Vladimir Dmitriev and Bill Halperin for useful discussions. The present work was supported by JSPS KAKENHI (Grant No.16K05444).

\end{document}